\documentclass{aa}  
\usepackage{caption}
\usepackage{bigstrut}
\usepackage{epstopdf}
\usepackage{epsfig}
\usepackage{graphicx}
\usepackage{txfonts}
\usepackage{natbib}
\usepackage{multirow}
\newcommand{\EG}{(CH$_2$OH)$_2$}
\newcommand{\MF}{HCOOCH$_3$}
\newcommand{\GLY}{CH$_2$OHCHO}

\begin{document}



   \title{Tentative detection of ethylene glycol toward W51/e2 and G34.3+0.2}

   \author{J.~M.~Lykke
          \inst{1},
          C.~Favre\inst{2},
          E.~A.~Bergin \inst{2}
          \and
          J.~K.~J\o rgensen \inst{1}
          }

   \institute{Centre for Star and Planet Formation, Niels Bohr Institute \& Natural History Museum of Denmark, University of Copenhagen, \O ster Voldgade 5-7, 1350 Copenhagen K., Denmark \\
              \email{jmlykke@nbi.ku.dk}
          \and
             Department of Astronomy, University of Michigan, 500 Church Street Ann Arbor, MI 48109, USA\\
             }

   \date{Received}


  \abstract
   {How complex organic -- and potentially prebiotic -- molecules are formed in regions of low- and high-mass star-formation remains a central question in astrochemistry. In particular, with just a few sources studied in detail, it is unclear what role environment plays in complex molecule formation. In this light, a comparison of relative abundances of related species between sources might be useful to explain observed differences.}
   {We seek to measure the relative abundance between three important complex organic molecules, ethylene glycol(\EG{}), glycolaldehyde (\GLY{}) and methyl formate (\MF{}), toward high-mass protostars and thereby provide additional constraints on their formation pathways.}
   {We use IRAM 30-m single dish observations of the three species toward two high-mass star-forming regions -- W51/e2 and G34.3+0.2 -- and report a tentative detection of \EG{} toward both sources.}
   {Assuming that \EG{}, \GLY{} and \MF{} spatially coexist, relative abundance ratios, HCOOCH$_3$/(CH$_2$OH)$_2$, of 31 and 35 are derived for G34.3+0.2 and W51/e2, respectively. \GLY{} is not detected, but the data provide lower limits to the HCOOCH$_{3}$/CH$_{2}$OHCHO abundance ratios of $\ge$193 for G34.3+0.2 and $\ge$550 for W51/e2. A comparison of these results to measurements from various sources in the literature indicates that the source luminosities may be correlated with the HCOOCH$_3$/(CH$_2$OH)$_2$ and HCOOCH$_3$/CH$_2$OHCHO ratios. This apparent correlation may be a consequence of the relative timescales each source spend at different temperatures-ranges in their evolution. Furthermore, we obtain lower limits to the ratio of (CH$_2$OH)$_2$/CH$_2$OHCHO for G34.3+0.2 ($\ge$ 6) and W51/e2 ($\ge$ 16). This result confirms that a high (CH$_2$OH)$_2$/CH$_2$OHCHO abundance ratio is not a specific property of comets, as previously speculated.}
   {}


   \maketitle

\section{Introduction}
\begin{table*}
 \caption[]{\label{tlines}Transitions of (CH$_2$OH)$_2$ in the observed frequency range.}
 \centering
 \small\addtolength{\tabcolsep}{-1pt} 
\begin{tabular}{lcccc}
 \hline \hline
  Transition &
Frequency &  
  E$_{up}$ &
  $\log_{10} (A_{ul})$&
  Rotational diagram \\
 & $[$MHz$]$ & $[$K$]$ & [s$^{-1}$] &  \\
 \hline
 \multicolumn{5}{ c }{3~mm observations} \bigstrut[t] \\
\hline
 $ 9 _{1 ,8 } \: v = 1   -  8_{ 1, 7} \: v =  0  $&  100333.6424 &  23.6 &-4.6773 & W51/e2 \\
$9 _{3 ,6 } \: v = 1   -  8 _{3, 5} \: v =  0  $&  100490.6121&  27.0 & -4.6800 &\\
$ 11 _{1,10}  \: v = 1  -  10_{ 2, 9 } \: v = 1 $&  102064.0770  &   33.9 & -4.6692& \\
$10 _{1,10}  \: v = 1  -   9 _{1 ,9} \: v =  0 $&  102291.5860  &    26.2 &-4.5798& \\
$ 9 _{2 ,7 } \: v = 1   -  8_{ 2 ,6} \: v =  0  $&  102539.4267 & 25.0 & -4.6355& \\
$ 10_{ 0,10}  \: v = 1  -   9_{ 0, 9} \: v =  0 $&  102689.8384 &  26.1 & -4.4419& \\
$11 _{9, 2}  \: v = 0  -  10 _{9, 1} \: v =  1 $&  105574.9490  &  72.2 & {-4.9342}& \\
$11 _{8, 4/3} \: v =  0  -  10_{ 8, 3/2 } \: v = 1 $&  {105644.1940}  & {   63.9} & {-4.5179}& \\
$ 11_{ 1,10}  \: v = 0  -  10_{ 1, 9 } \: v = 1 $&  {105701.0020}  &  {  33.6} & {-4.3644}& \\
$ 11 _{7, 5/4}  \: v = 0  -  10_{ 7, 4/3 } \: v = 1 $&  {105735.3520}  & {   56.5} & {-4.4051}& \\
$ 11 _{3, 9} \: v =  0  -  10 _{3, 8} \: v =  1 $&  {105834.6960}  &   { 36.9} & {-4.5676}& \\
$ 11 _{6, 6/5} \: v =  0  -  10_{ 6, 5/4} \: v =  1  $&  {105866.9546}  & {   50.1} &   {-4.3228}& \\
$ 11 _{5, 7}  \: v = 0  -  10_{ 5, 6} \: v =  1 $&  {106072.0524} & {   44.7} & {-4.6197}& \\
$ 11 _{5, 6} \: v =  0  -  10_{ 5, 5} \: v =  1 $& {106088.9178} & {44.8} & {-4.5104} &\\
 \hline
\multicolumn{5}{ c }{1~mm observations}  \bigstrut[t] \\
\hline
 $24_{17,8/7} \: v= 0- 23_{17, 7/6} \:v=1$ & 238828.3252	&  289.3 & -3.4806   &   \\
 $24_{16,9/8} \:v= 0- 23_{16,8/7} \:v= 1 	 $ & 238904.3833	&  273.1& -3.4330  &   \\
 $24_{15,10/9} \: v= 0 -  23_{15, 9/8} \: v= 1    $ & 238993.6343	&  257.9 & -3.3925   &  \\
 $30_{12,19/18} \: v= 1 - 30_{11,19/20} \: v= 0  $ &  238994.0910   &  299.7    & -4.8643  &  \\
 $22 _{5,17} \: v= 1 -     21_{5,16} \: v= 0    	 $ &   238994.7545	    &  138.6 & -3.5149   &  \\
 $24_{13,12/11} \: v= 0-  23_{13,11/10} \: v= 1  $ &	239231.3329 &  230.4 & -3.3269 	&  \\
 $ 24_{12,13/12} \: v= 0-  23_{12,12/11} \: v=1  $	 & 239395.0677	&  218.2 & -3.3002  & \\
 $ 24_{11,14/13} \: v= 0-  23_{11,13/12} \: v= 1 $ &	239605.3330	&  206.9 & -3.2765  &  \\
 $ 25 _{3,23} \: v= 0-   24_{ 3,22} \: v= 1 	     $     & 239792.7984	& 161.6 & -3.4697 &  \\
 $  24_{10,15/14} \: v= 0-23_{10,14/13} \: v= 1  $ & 239883.5631	&  196.7 & -3.2555  & W51/e2  \\
 $ 25_{ 2,23} \: v= 0 -  24_{ 2,22} \: v= 1 	     $     & 239957.1582	& 161.6 & -3.4917  &  \\
 $ 23_{ 4,20} \: v= 1- 22_{ 4,19} \: v= 0 	         $ & 239980.0390	&  144.1 & -3.4495 &   \\
 $ 24_{3,21} \: v= 0  - 23_{ 3,20} \: v= 1 	         $ & 240147.9627	&  155.2 & -3.4704  & \\
 $  24_{ 9,16/15} \:v= 0 -23_{ 9,15/14} \: v= 1 $ & 240265.7687	&	 187.4 & -3.2365   &	 \\
 $ 22 _{4,19} \: v= 0 -  21_{3,18} \: v= 0 	         $ & 240608.1463	&  132.6 & -3.9839  & \\
 $ 25_{ 1/0,25} \: v= 1 - 24_{ 1/0,24} \: v= 0 	 $ & 240778.3210	&	 148.0	 & -3.1643 &  \\
 $ 24_{8,17} \: v= 0 -  23_{ 8,16} \: v= 1 	         $ & 240807.8795	&  179.2 & -3.5201 & W51/e2 and G34.3+0.2 \\
 $ 24_{ 8,16} \: v= 0  -  23_{ 8,15} \: v= 1  	     $     & 240828.8863	&  179.2 & -3.5199  &  \\
 $ 22_{ 4,18} \: v= 1  - 21_{ 4,17} \: v= 0  	         $ & 240875.3680	&  135.5 & -3.4905  &  \\
 $ 59_{10,50} \: v= 1  -  59_{ 9,51} \: v= 1  	     $     & 240888.3533	& 925.5 & -4.4992   &  \\
 $ 24_{ 5,20} \: v= 0     23_{ 5,19} \: v= 1  	     $     & 241291.2695	& 160.7  & -3.5046  &  \\
 $ 24_{ 7,18} \: v= 0 -  23_{ 7,17} \: v= 1   	     $     & 241545.2626	&  172.1 & -3.5045   &  \\
 $ 24 _{7,17} \: v= 0  -  23_{ 7,16} \: v= 1  	     $     & 241817.6908	& 172.1  & -3.5026   &  \\
 $ 24_{ 6,19} \: v= 0   - 23_{ 6,18} \: v= 1  	     $     & 241860.7330	& 166.0 & -3.5459   &  \\
 $ 23_{15,9/8} \: v= 1 - 22_{15, 8/7 } \: v= 0 	 $ & 242244.6928	& 246.4 & -3.4004 &  \\
 $	  23_{14,10/9 } \: v= 1 - 22_{14, 9/8 } \: v= 0  $ &242246.3392	&	232.2  & -3.3608 & \\
 $ 23_{16, 8/7 } \: v= 1  -  22_{16, 7/6 } \: v= 0   $	 & 242267.1576	& 261.6 & -3.4470 	& \\
 $	 23_{13,11/10 } \: v= 1  -22_{13,10/9 } \: v= 0  $ &242277.7167	&	218.9  & -3.3268  &   \\
 $ 23_{17, 7/6 } \: v= 1  -  22_{17, 6/5 } \: v= 0   $	 & 242309.7823	& 277.8 & -3.5027 	&  \\
 $ 23_{12,12/11 } \: v= 1  - 22_{12,11/10 } \: v= 0  $ &	242347.0348 & 206.7 & -3.2973 &\\
 $	 23_{18, 6/5 } \: v= 1  - 22_{18, 5/4 } \: v= 0  $ &	242369.7298 &	 295.0 & -3.5709  &  \\
 $ 26_{ 2/1 ,25} \: v= 0  -  25_{ 2/1 ,24} \: v= 1   $	 & 242389.2838	&  166.9 & -3.1601  & \\
 $	 23_{11,13/12 } \: v= 1 -22_{11,12/11 } \: v= 0  $  &242466.6909	&	 195.4 & -3.2714  &   \\
 $  23_{10,14/13 } \: v= 1 -22_{10,13/12 } \: v= 0   $  & 242656.2283	&  185.2 & -3.2486  & W51/e2 and G34.3+0.2 \\
$23_{ 3,20} \: v = 1   - 22 319 \: v = 0$ & 242897.4293 & 143.6 & -3.7464 & \\ 
 $ 23_{ 9,15/14 } \: v=1 -22_{ 9,14/13 } \: v= 0     $  &	242948.2909 &	175.9 & -3.2282  & W51/e2 \\
 $ 24_{ 6,18} \: v= 0  -  23_{6,17} \: v= 1            $	&          243259.7398	& 166.3   & -3.8001 &                                  \\
 $ 23_{8,16} \: v= 1  - 22_{ 8,15} \: v= 0  	         $  &          243396.0258	& 167.7  & -3.5106 &                                     \\
 $ 23 _{8,15} \: v= 1  - 22_{ 8,14} \: v= 0            $	&          243408.4357	& 167.7 & -3.5105 &                                    \\
 $ 23 _{5,19} \: v= 1  - 22 _{5_18} \: v= 0        	 $ &          243636.5668	& 149.1 & -3.4704  &                                      \\
$23_{ 7,17} \: v = 1 -   22_{ 7,16} v = 0$ & 244054.2878 & 160.5 & -3.4934 & \\
 $ 23_{7,16} \: v= 1  - 22_{ 7,15} \: v= 0         	 $ &          244233.4545	& 160.5 & -3.4927 &                                       \\
 $ 24_{ 3,22} \: v= 1  - 23_{ 3,21} \: v= 0       	 $ &          244399.6722	& 150.1 & -3.4754  &                                         \\
 $ 24 _{2,22} \: v= 1  -  23_{ 2,21} \: v= 0           $	&          244685.1477	& 150.1 & -3.4399 &                                          \\
 $ 23 _{6,18} \: v= 1  -  22_{6,17} \: v= 0 	         $ &          244879.9193	& 154.4  & -3.5188   &                                          \\
 $  27 _{1/0 ,27} \: v= 0  - 26_{ 1/0 ,26} \: v= 1   $ &    245022.7624	&	171.4 & -3.1407 &                                   \\ 
 $ 24_{ 6,19} \: v = 1 -   23_{ 6,17} \: v = 1 $ & 245410.9218 & 166.3 & -3.7580 & \\
\hline

\end{tabular}
\tablefoot{Catalog values for the \EG{} transitions that can be excited in the observed frequency range for W51/e2 and G34.3+0.2. Some faint transitions have been excluded from this table, please section \ref{sec:resandan}. The source names are listed for the lines that are used in the rotational diagram analysis.}
\end{table*}

A central question in astrochemistry is how, where, and when complex organic molecules form. Although the total number of detected molecules in the interstellar medium\footnote{http://www.astro.uni-koeln.de/cdms/molecules} (ISM) continuously increase \citep[e.g.][]{Herbst2009}, the answers to the question above still remain unclear. To date there is no consensus on how complex organic molecules form in dense regions of the ISM, despite the increasing number of detections. One promising suggestion was that warm gas-phase chemistry, following evaporation of simple ices, could be a primary formation pathway \citep[e.g.][]{Millar1991,Charnley1992}. However, more recent laboratory experiments and chemical modeling has shown that this mechanism is not effective enough to account for the observed abundances \citep[e.g.][]{Geppert2006}. An alternative formation mechanism involves UV induces radicals. \cite{Garrod2008} propose that radicals, during the warm up phase, can migrate on the grains surfaces and form complex species which are then released into the gas-phase at higher temperatures. The initial ice composition and the amount of UV radiation have also proved to play an important part of the formation process \citep{Oberg2009}. In order to determine the formation pathway of COMs, it is useful to determine abundance ratios between different related species as these ratios, in comparison with the predicted abundance ratios from chemical models, can provide constraints on the formation processes. Indeed, variations in the abundance profiles can reflect the physical and chemical conditions that are occurring. It is therefore important to observe these species in different environments. 

Some of the simplest species in this context include the oxygen-bearing complex organic molecules associated to glycolaldehyde, $\text{CH}_2 \text{OHCHO}$, including its isomer methyl formate (HCOOCH$_3$ or CH$_{3}$OCHO) and the reduced alcohol version of \GLY{}, ethylene glycol, $(\text{CH}_2 \text{OH})_2$ (also commonly known as anti-freeze). By constraining the relative abundances of these species in different environments the hope is to be able to explore, e.g., the importance of initial chemical conditions, temperature and irradiation in the formation for comparison, e.g., to laboratory experiments \citep[e.g.][]{Oberg2009} and as input for sophisticated chemical models \citep[e.g.,][]{Garrod2008}. For example, in their laboratory experiments \cite{Oberg2009} show that the relative abundances of \MF{} to \GLY{} and \EG{} are strongly dependent on both the ice temperature and exact ice composition in terms of the relative amounts of CO and CH$_3$OH.

So far, \EG{} has been detected toward high-mass sources such as the Galactic center source Sgr B2(N) by \citet{Hollis2002}, marginally toward W51 e1/e2 by \citet{Kalenskii2010} and recently also toward the Orion Kleinmann-Low nebula by \citet{Brouillet2015}. \EG{} has also been observed toward the low-mass Class 0 protostars IRAS 16293--2422 \citep[][in prep.]{Jorgensen2012} and NGC 1333 IRAS 2A \citep{Maury2014,Coutens2015} as well as toward the intermediate-mass protostar NGC 7129 FIRS2 \citep{Fuente2014}. \GLY{} has been detected toward Sgr B2(N) (\citealt{Hollis2000, Hollis2001,Hollis2004}; \citealt{Halfen2006}; \citealt{Requena2008}), the high-mass hot molecular core G31.41+0.31 \citep{Beltran2009}, in IRAS 16293--2422 by \citet{Jorgensen2012} and recently also in NGC 7129 FIRS2 \citep{Fuente2014}, NGC 1333 IRAS 2A \citep{Coutens2015,Taquet2015} and NGC 1333 IRAS 4A \citep{Taquet2015}. \MF{} is the most abundant isomer of \GLY{} and has previously been observed in numerous hot cores and corinos, (e.g. \citealt{Blake:1987}; \citealt{Bisschop2007}; \citealt{Demyk:2008}; \citealt{favre:2014a,Favre2011}). There are some notable differences in terms of the abundance ratios. For example, \cite{Coutens2015} find a (CH$_2$OH)$_2$/CH$_2$OHCHO ratio of $\sim$5 in NGC 1333 IRAS 2A, while Jørgensen et al. (2012 and in prep.) find a lower value of $\sim$1 in IRAS 16293-2422. These changes hint that it might be useful to explore these ratios in different sources sampling similar physical conditions.

This paper presents IRAM 30--m observations of \EG{} and \GLY{} toward the high mass protostars W51/e2  \citep[distance = 5.4 kpc,][]{Sato2010} and G34.3+0.2 \citep[distance = 3.8 kpc,][]{Kurtz2000}. The aim of this study is to determine the relative abundance of \EG{} and \GLY{} to \MF{} and compare to other sources from the literature. Overall our aim is to investigate the use of abundance ratios of species that are believed to be chemically related to explore the origin of complex molecules in the dense interstellar medium. In Section~\ref{sec:obs}, the IRAM 30--m observations are presented. Data modeling, results and analysis are presented in Section~\ref{sec:resandan} and discussed in Section~\ref{sec:disc}.

\section{Observations}
\label{sec:obs}
\begin{table*}
\caption{Transitions of (CH$_2$OH)$_2$ observed toward W51/e2 and G34.3+0.2 with the IRAM 30--m telescope.}
\label{t_w51_obsogt_g34}
\centering
\small\addtolength{\tabcolsep}{-1pt} 
\begin{tabular}{lccccccc}
\hline\hline
Transition  & Freq. & E$_{up}$ & log$_{10}$A$_{ul}$ & ${\int{T_{mb} \Delta v }}$ & V$_{LSR}$ & $\Delta v$ & Peak T$_{mb}$ \bigstrut[t] \\
$J^ {\prime} _{K_a,K_c} \: v^{\prime} - J^{\prime \prime}  _{K_a,K_c} \: v^{ \prime \prime}$        & $[$MHz$]$ & $[$K$]$ & $[$s$^{-1}]$ & $[$ K km s$^{-1}]$ & $[$km s$^{-1}]$ & $[$km s$^{-1}]$ & $[$K$]$ \\
\hline
\multicolumn{8}{ c }{W51/e2} \bigstrut[t] \\
\hline
 $9 _{1 ,8 } \: v = 1   -  8_{ 1, 7} \: v =  0  $&  100333.64 &  24 & -4.6773 & 0.2 & 55.1 & 5.5 & 0.03  \bigstrut[t]  \\
$  24_{10,15/14} \: v= 0-23_{10,14/13} \: v= 1  $ & 239883.56	&  197 & -3.2555 & 2.7 & 55.6 & 6.8 & 0.38\\
$ 24_{8,17} \: v= 0 -  23_{ 8,16} \: v= 1 $ & 240807.88	&  179 & -3.5201 & 2.0 & 56.7 & 5.9 & 0.32 \\
$23_{10,14/13} \: v = 1 -   22_{10,13/12} \: v = 0$ & 242656.23 & 185 & -3.5496 & 4.0 & 56.9 & 7.4 & 0.50 \\

$23_{9,15/14 } \: v = 1   -   22_{9,14/13 } \: v =  0 $ & 242948.29 & 176 & -3.5292 & 1.4 & 56.5 & 5.0 & 0.27 \\

\hline
\multicolumn{8}{ c }{G34.3+0.2}  \bigstrut[t] \\
\hline
$ 24_{8,17} \: v= 0 -  23_{ 8,16} \: v= 1 $ & 240807.88	&  179 & -3.5201 & 1.0 & 57.7 & 4.1 & 0.23 \\
$23_{10,14/13} \: v = 1 -   22_{10,13/12} \: v = 0$ & 242656.23 & 185 & -3.5496 & 1.0 & 58.4 & 5.0 & 0.19 \\
\hline
\end{tabular}
\tablefoot{The table lists spectroscopic parameters and observed quantities from Gaussian fits to the detected \EG{} lines: integrated line intensity ($\int{T_\mathrm{mb}} \Delta v$), line position (V$_{LSR}$), line width ($\Delta v$) and peak temperature (Peak $T_{mb}$). The error of $\int{T_\mathrm{mb}} \Delta v$ is estimated to $\sim 30 \%$ as determined by the observational uncertainty, while the errors on V$_{LSR}$ and $\Delta v$ are $\sim$ 1.0 and $\sim$ 2.0 km s$^{-1}$, respectively. The line intensities have been added together for the hyper fine structure transitions (denoted by $/$ in the quantum numbers).}
\end{table*}

The observations were performed with the IRAM 30--m telescope at Pico Veleta, Spain on December 13 and 14, 2012 for W51/e2 and G34.3+0.2, respectively. The coordinates of the phase tracking center used for the two sources were $\alpha_{J2000}$ = 19$^{h}$23$^{m}$43$\fs$9, $\delta_{J2000}$ = 14$\degr$30$\arcmin$34$\farcs$8 for W51/e2 and $\alpha_{J2000}$ = 18$^{h}$53$^{m}$18$\fs$6, $\delta_{J2000}$ = 01$\degr$14$\arcmin$58$\farcs$0 for G34.3+0.2. The observations were performed in position switching mode using $[-600 \arcsec,0 \arcsec]$  as reference for the OFF positions. The spectral setup was chosen to target \EG{} and \GLY{} in the observations. The EMIR receiver was used in dual band polarisation (E090/E230) in connection with \textit{i)} the 200~kHz Fourier transform spectrometer (FTS) back-end in the frequency ranges 99.70--106.3~GHz and 238.2--246.0~GHz for the E090 and E230 bands, respectively; \textit{ii)} the WILMA back-end in the frequency ranges 99.88--103.6~GHz and 238.4--242.1~GHz for the E090 and E230 bands, respectively. Data reduction was performed using The Continuum and Line Analysis Single-dish Software\footnote{http://www.iram.fr/IRAMFR/GILDAS} (CLASS). The resulting FTS spectra were smoothed to a spectral resolution of 1.15 km~s$^{-1}$ for the E090 band and 1.22 km~s$^{-1}$ for the E230 band while the WILMA spectra were smoothed to 5.88 km~s$^{-1}$ and 2.49 km~s$^{-1}$ for the E090 and E230 bands, respectively. Spectra resulting from the FTS back-end presented a standing wave pattern and the WILMA observations were therefore used as a sanity check for the FTS data and to confirm the detections made in the FTS observations for the lines where the frequency of the two sets of observations match. The standing wave present in the FTS observations was removed using a fast Fourier transform in the data reduction.

Throughout this paper, the intensity is given as the main beam brightness temperature ($T_\mathrm{mb}$), which is defined as 
\[T_\mathrm{mb} = T_\mathrm{a} \times \frac{F_\mathrm{eff}}{B_\mathrm{eff}} , \]
where $T_\mathrm{a}$ is the antenna temperature, $F_\mathrm{eff}$ is the forward efficiency and $B_\mathrm{eff}$ is the beam efficiency. The values used were $F_\mathrm{eff} = 0.94$ and 0.92 and $B_\mathrm{eff}$ = 0.78 and 0.58 for 1~mm and 3~mm respectively. The half-power beam sizes are $\sim 10  \arcsec $ and $\sim 24 \arcsec $  for the observations at 1~mm and 3~mm, respectively. The reduced data-sets are available for download at http://youngstars.nbi.dk/projects/HighMass\_organics/index.html. 

\section{Analysis and Results}
\label{sec:resandan}
Spectra from both sources show a rich forest of lines characteristic for high-mass sources. A total of 21 and 19 lines have peak temperature above 5~K for W51/e2 and G34.3+0.2, respectively; the strongest lines in both spectra is the CS 5--4 transition at 244935 MHz. It is important to know that there is still a remnant oscillation in the baseline even after the removal of the standing wave in the spectra. This increases the RMS noise of the spectra, which complicates the analysis of faint lines. Therefore, in addition to the global baseline subtraction, an additional local zero or first order baseline subtraction was therefore performed before making a Gaussian fit of the \EG{} lines. This additional baseline subtraction is applied in a range of $\pm$ 100~km~s$^{-1}$ for each line. As for the data reduction, CLASS was used for the additional baseline subtraction as well as Gaussian profile fits to the detected lines. The resulting RMS of the local baseline is $\sim 60$~mK for the 1~mm observations and $\sim 7$~mK for the 3~mm observations.

In order to ensure proper line identification, we have checked the line observations against entries in Splatalogue database for astronomical spectroscopy\footnote{www.splatalogue.net}. In addition, we have made a reference model where we produced synthetic spectra for the line emission of common species (CH$_3$OCH$_3$, HCOOH, CH$_3$CHO, CH$_3$OH, C$_2$H$_5$OH and CH$_3$CN) in order to visually exclude lines that are blended with any of these molecules. Table \ref{tlines} lists spectroscopic parameters for the \EG{} transitions that can be excited in the observed frequency range. Transitions with $\log (A_\mathrm{ul}) < - 5$ for the 3~mm observations and $\log (A_\mathrm{ul}) < - 4$ and $E_\mathrm{up} > 300$~K for the 1~mm observations have been excluded from the table, as these transitions are predicted from the synthetic spectra to have peak intensities $ \la 0.02$~K, i.e., not detectable.

The spectroscopic data for \EG{} come from \citet{Christen:2003} and \citet{Christen:1995} and are available from the CDMS database\footnote{http://www.astro.uni-koeln.de/cdms} \citep{Muller2001,Muller2005}, while the spectroscopic data for \MF{} and \GLY{} are from \citet{Ilyushin:2009} and \citet{Carroll2010} respectively, available from the JPL database\footnote{http://spec.jpl.nasa.gov/home.html} \citep{Pickett:1998}. For the analysis, the \EG{} lines that are reasonably well separated and have a peak temperature above 3~$\sigma$ were selected. For W51/e2 we have included the lines at $\sim 100333$ and $\sim 242656$~MHz although they are partially blended, as it is possible to distinguish the \EG{} peak from the other peaks and perform a multiple Gaussian fit which includes all relevant peaks.

\begin{figure*}
\includegraphics[width=18cm]{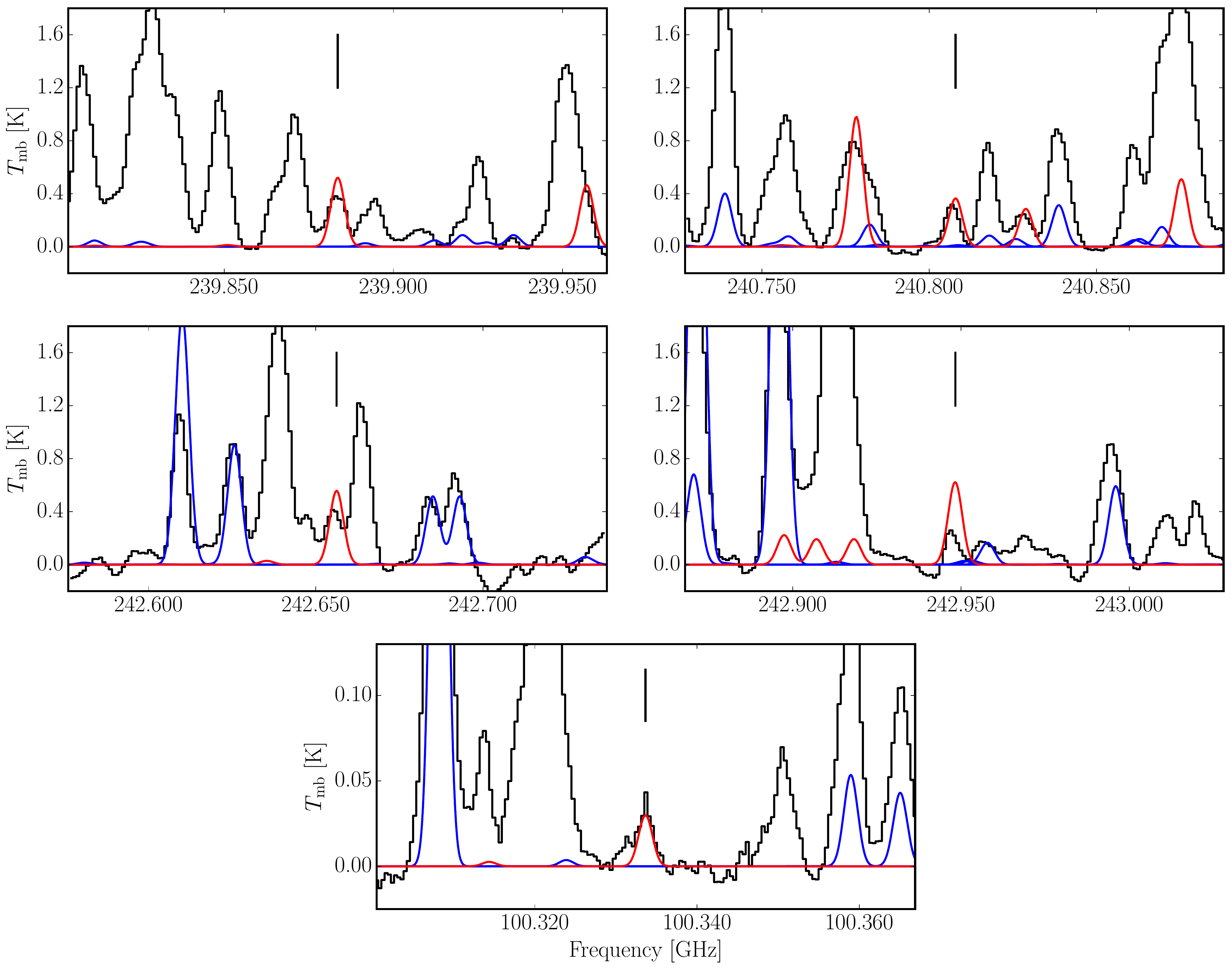}
\caption{Observed (black) and synthetic (red) spectra of the five transitions of \EG{} in W51/e2 which are used for the analysis. Also shown are the synthetic spectra of all other species investigated in this study (blue) to demonstrate that the \EG{} lines in question are not blended. The transitions are located at 239883.56, 240807.88, 242656.23, 242948.29 and 100333.64~MHz -- marked by black vertical lines in the plot. The displayed frequency region in each plot corresponds to~$\sim$~200~km~s$^{-1}$.}
\label{fig:zoom_lines_w51}
\end{figure*}

\begin{figure*}
\includegraphics[width=18cm]{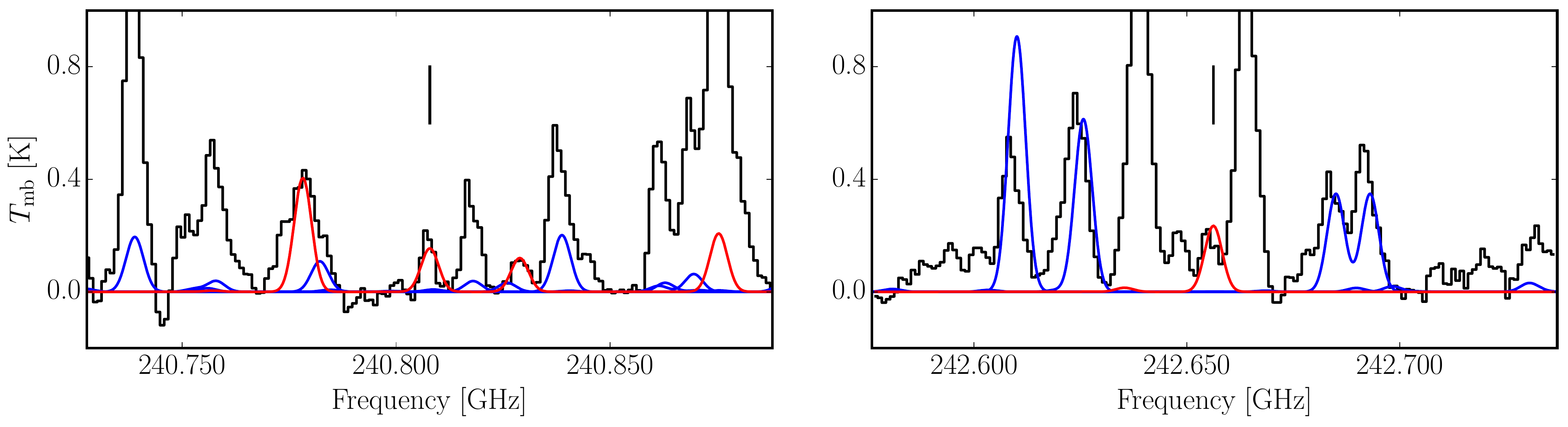} 
\caption{Same as for Figure \ref{fig:zoom_lines_w51} but for G34.3+0.2. The transitions are located at 240807.88 and 242656.23~MHz -- marked by black vertical lines in the plot.}
\label{fig:zoom_lines_g34}
\end{figure*}

Figure~\ref{fig:zoom_lines_w51} and \ref{fig:zoom_lines_g34} show a zoom-in of the area of $\sim$ 200~km~s$^{-1}$, which corresponds to $\sim$ 160~MHz in the 1~mm observations and $\sim 70$~MHz in the 3~mm~observations, around each detected line in both sources after the local baseline subtraction. Superimposed onto the observed spectra are the synthetic spectra of \EG{} as well as the other species investigated here, as to demonstrate that the chosen \EG{} lines are not blended. The fit to the entire observed spectra for both sources are shown in Figure \ref{fig:long_spec_w51_1}-\ref{fig:long_spec_g34_3mm_2} in the appendix. Table \ref{t_other_coms} in the appendix lists the estimated column densities of the molecules in the reference model. Table \ref{t_w51_obsogt_g34} lists spectroscopic parameters and the observed quantities from the fits of the \EG{} lines: the integrated line intensities ($\int{T_\mathrm{mb}} \Delta v$), position (V$_\mathrm{LSR}$), line width ($\Delta v$) and peak temperature (Peak $T_\mathrm{mb}$).

A total of 35 and 52 well separated lines have been detected above the 3~$\sigma$ level in the 1~mm and 3~mm data for \MF{} in W51/e2 and G34.3+0.2 respectively. For \EG{}, 4 and 2 lines, toward W51/e2 and G34.3+0.2 respectively, are detected in the 1~mm observations; while only one line is detected at 3~mm toward for W51/e2. Many other \EG{} lines as well as some potential \GLY{} lines are present in both data-sets, but they are either blended with other species or too faint to be properly detected above a 3~$\sigma$ limit.

\subsection{Modeling}
\label{subsec:model}
Assuming LTE and optically thin emission, the rotational diagram method can be used to determine the column density, $N_\mathrm{tot}$, and the rotational temperature, $T_\mathrm{rot}$ \citep{Goldsmith1999}. The approach used in this work follows the formalism described by \citet{Goldsmith1999} with the following assumptions:
\begin{enumerate}
\item \EG{}, \GLY{} and \MF{} are emitting from the same region and thus have the same source size, $\theta_\mathrm{source}$.
\item \EG{}, \GLY{} and \MF{} are in LTE, which implies that the excitation temperature, $T_\mathrm{ex}$, is equal to the kinematic temperature $T_\mathrm{kin}$, for all three species. Using the rotational diagram method one obtains the rotational temperature which in LTE is $T_\mathrm{kin}$ = $T_\mathrm{rot}$ = $T_\mathrm{ex}$.
\item The observations at 1~mm and 3~mm trace the same gas.
\end{enumerate}
According to our assumptions, the source size and $T_\mathrm{rot}$ derived from the \MF{} analysis can be used to derive the column density of \EG{} and set an upper limit on the column density of \GLY{}. A possible caveat is that the large beam size of our observations compared to the source sizes is not guaranteeing that the measured lines arise from species present in the same gas. Specifically, we obtain similar mean values of V$_\mathrm{LSR}$ for \EG{} and \MF{} in both sources ($56.4 \pm 0.5$~km~s$^{-1}$ and $56.0 \pm 0.8$~km~s$^{-1}$ in W51/e2 along with $57.7 \pm 0.5$~km~s$^{-1}$ and $58.4 \pm 1.2$~km~s$^{-1}$ in G34.3+0.2 for \EG{} and \MF{}, respectively) so here the assumption appears reasonable.

The same formalism in \cite{Goldsmith1999} used to create a rotational diagram can also the used to generate a synthetic spectrum of the line emission of a specific molecules by using the source size, $\theta_\mathrm{source}$, the rotational temperature, $T_\mathrm{rot}$, the line width, $\Delta v$, and the column density, $N_\mathrm{tot}$, as input parameters. It is possible to correct for the optical depth of the line emission in case it deviates from being completely optically thin \citep{Goldsmith1999}. A synthetic spectrum generated with inputs parameters derived from the rotational diagram therefore serves as a check for self-consistency of the result. We use adopted source sizes from the literature and determine the rotational temperature and column densities in the following analysis.

The observed width of the detected lines have been determined from the Gaussian fit. For \MF{} in W51/e2 we obtain mean values of $\Delta v = 6.4 \pm 0.8$ km s$^{-1}$ in the 1~mm data and $\Delta v = 6.1 \pm 1.1$ km s$^{-1}$ in the 3~mm data. In the same source, the mean value of the line width of the \EG{} lines is $6.3 \pm 0.9$ km s$^{-1}$ for the 1~mm data and the only detected value is 5.5  km s$^{-1}$ in the 3~mm data. For G34.3+0.2 we find $\Delta v = 6.5 \pm 1.5$ km s$^{-1}$ in the 1~mm data and $\Delta v = 7.3 \pm 1.3$ km s$^{-1}$ in the 3~mm data for \MF{} and $\Delta v = 4.5 \pm 0.5$ km s$^{-1}$ for the \EG{} lines in the 1~mm data. In the below analysis, a fixed line width of 6.0 km s$^{-1}$ was chosen as input parameter in synthetic spectra for all species.  
\begin{table}[h!]
\caption{Input parameters for synthetic spectra}
\label{tresults}
\centering
\begin{tabular}{lcc}
  \hline\hline
  Parameter & G34.3+0.2 & W51/e2 \bigstrut[t] \\
 \hline
$\theta_\mathrm{source} [\arcsec]$ & 7.6 & 2.4 \bigstrut[t] \\
$T_\mathrm{rot}$[K] & 140 & 120  \\
$\Delta v$ $[\mathrm{km s}^{-1}]$& 6.0 & 6.0 \\
HCOOCH$_3$ $N_\mathrm{tot} [\mathrm{cm}^{-2}]$ & $5.8 \times 10^{16}$ & $1.1 \times 10^{18}$  \\
(CH$_2$OH)$_2$ $N_\mathrm{tot} [\mathrm{cm}^{-2}]$ & $1.9 \times 10^{15}$ & $3.1 \times 10^{16}$  \\
CH$_2$OHCHO $N_\mathrm{tot}[\mathrm{cm}^{-2}]$ & $ \le 3 \times 10^{14}$ & $\le 2 \times 10^{15}$ \\
\hline
\end{tabular}
\tablefoot{The source sizes, rotational temperatures, line widths and column densities which are used as input parameters when creating synthetic spectra. As described in the text, the estimated uncertainties are $\pm 40$~K for $T_\mathrm{rot}$ and $\pm 30$--40$\%$ for the column density.}
\end{table}

\subsection{\MF{}}
In the rotational diagram analysis for W51/e2, a source size of 2.4~$\arcsec$ \citep{Zhang1998} is applied and results in a rotation temperature of $T_{\rm rot} = 120$~K and a column density of $1.1 \times 10^{18}$~cm$^{-2}$. We estimate a 40 K uncertainty on the rotational temperature, which contribute to a $\sim 20 \%$ error on the column density. This, combined with an observational uncertainty of $\sim 30 \%$, returns an estimated overall error of 30 --40~\% for the column density. For G34.3+0.2 a source size of 7.6~$\arcsec$ \citep{Remijan2003} is applied and results in $N_\mathrm{tot} = 5.8 \times 10^{16}$ cm$^{-2}$, and $T_\mathrm{rot} = 140$~K for \MF{}. As in the case for W51/e2, the estimated uncertainty of the column density is 30 -- 40\% and 40~K for the temperature. Results from both sources are given in Table \ref{tresults} and we have checked these results by using the parameters derived from the rotation diagram.

\subsection{\EG{}}
Numerous \EG{} lines are blended with other species or have line intensities below 3~$\sigma$. A total of two and five reasonable well separated lines above the 3~$\sigma$ limit toward G34.3+0.2 and W51/e2 respectively were chosen for this analysis (see Table~\ref{t_w51_obsogt_g34} and \ref{tlines} for the spectroscopic data and Figure~\ref{fig:zoom_lines_w51} and~\ref{fig:zoom_lines_g34} for the zoom-in of the spectra around the lines). With only 2--5 unblended lines, the assignment of \EG{} cannot be considered as a firm detection. However, even if our tentative detection of \EG{} is not confirmed, this measurement represents a useful upper limit for the column density of \EG{}, which still can provide important information when compared to the \MF{} detection.

When only a few data points are available, a statistically reliable result of the rotational temperature cannot be obtained from the traditional use of the rotational diagram method. Thus, we assume that \EG{} and \MF{} emit at the same T$_\mathrm{ex}$ which in LTE is equal to T$_\mathrm{rot}$. The uncertainty of the column density is, as for \MF{}, estimated to 30--40$\%$.

Using $\theta_\mathrm{source} = 7.6\arcsec$ and $T_\mathrm{rot}=140$~K, a column density of $1.9 \times 10^{15}$~cm$^{-2}$ is derived for \EG{} toward G34.3+0.2. For W51/e2, a source size of $2.4 \arcsec$ and $T_\mathrm{rot} = 120$~K returns N$_\mathrm{tot} = 3.1 \times 10^{16}$~cm$^{-2}$ for \EG{}. These results, listed in Table \ref{tresults}, are used as input parameters for synthetic spectra in order to perform a sanity check. We checked each line in the synthetic spectra against the observed spectra for each source and they all seem to provide a reasonable match, at least within the estimated uncertainty. Figure \ref{fig:zoom_lines_w51} and \ref{fig:zoom_lines_g34} shows the comparison of the synthetic spectra against the observed spectra. It is evident from Figure \ref{fig:zoom_lines_w51} that the synthetic spectrum in the 1~mm data toward W51/e2 slightly overproduces the observed spectrum, while it reproduces the observed line in the 3~mm data. As this line has a lower E$_\mathrm{up}$ this could indicate that a lower excitation temperature would give a better fit. Indeed a fixed T$_\mathrm{rot} = 70$~K in the rotational diagram returns the same column density, N$_\mathrm{tot}=3.1 \times 10^{16}$cm$^{-2}$. 
 
\subsection{\GLY{}}
Several lines in the data might be assigned to \GLY{}. However, they are either blended with the emission of another molecules or too faint (i.e. below the 3~$\sigma$ limit), thus we cannot claim a detection. Nevertheless, we are able to obtain an upper limit for the column density. Following the assumptions stated in section \ref{subsec:model}, synthetic spectra were generated using $\theta_\mathrm{source}$, $T_\mathrm{rot}$ and $\Delta v$ in Table \ref{tresults}, allowing $N_\mathrm{tot}$ to vary. More specifically, the upper limit of the column density is then determined by increasing the value until the synthetic spectra would overproduce the observed spectra intensities at the locations of the \GLY{} lines in the observed spectra. The upper limit on the column density is $ \le 3 \times 10^{14} \mathrm{cm}^{-2}$ toward G34.3+0.2 and $ \le 2 \times 10^{15} \mathrm{cm}^{-2}$ toward W51/e2. The uncertainty of the column density is, as for \MF{}, estimated to 30--40$\%$.

\subsection{Relative abundances}
Following the three assumptions listed in in section \ref{subsec:model}, it is possible to calculate the relative abundance of the species at the rotational temperature and source size found for each source. The following abundance ratios have been computed: HCOOCH$_3$/(CH$_2$OH)$_2$, HCOOCH$_3$/CH$_2$OHCHO and (CH$_2$OH)$_2$/ CH$_2$OHCHO. As \GLY{} is not detected, it is only possible to set an upper limit on the column density. The HCOOCH$_3$/CH$_2$OHCHO and (CH$_2$OH)$_2$/ CH$_2$OHCHO ratios are therefore \textit{lower limits}. All three relative abundance ratios for W51/e2 and G34.3+0.2 are listed in Table~\ref{tcompare} along with previous measurements toward high-mass stars-forming regions, a hot core, an intermediate-mass protostar, low-mass protostars, molecular clouds toward the Galactic Centre and comets.
\begin{table*}
 \caption[]{\label{tcompare}Relative abundances of \EG{}, \MF{} and \GLY{} in different sources}
 \centering
\begin{tabular}{lccccc}
 \hline \hline
source &  
  HCOOCH$_3$/(CH$_2$OH)$_2$ &
  HCOOCH$_3$/CH$_2$OHCHO &
  (CH$_2$OH)$_2$/ CH$_2$OHCHO&
  $L_\mathrm{bol}[L_{\sun}]$ &
  References  \bigstrut[t] \\
 \hline
W51/e2 (this study) & 35 &  $>$ 550 & $>$ 16 & $4.7 \times 10^6$ & 1,2 \\
Orion-KL$^{(a)}$   & 35  & > 200-300 & $>$ 12 & $10^5$ & 3,4 \bigstrut[t] \\
G34.3+0.2 (this study) & 31 & $>$ 193 & $>$ 6 &$2.8 \times 10^5$ & 1,5 \\
Sgr B2(N) LMH$^{(b)}$ & -- & 52 & -- & $10^7$ & 6,7\\
G31.41 + 0.31$^{(c)}$ & -- & < 34 & -- & $1.8 \times 10^5$ & 8,5 \\
NGC 7129 FIRS2$^{(d)}$ & $\sim$ 15 & $\sim$ 30 & $\sim$ 2 & 500 & 9 \\
IRAS NGC 1333 2A$^{(e)}$ & 4 & 20 & 5 & 20 & 10,11 \\
IRAS 16293--2422$^{(e)}$ & $\sim$ 13 & $\sim$ 13 &  $\sim$ 1 & 27 & 12,13 \\
IRAS NGC 1333 4A$^{(e)}$ & -- & 10 & -- & 7.7 & 14\\
MC G + 0.693$^{(f)}$  & 4.3    & 5.2  & 1.2 & -- & 15 \\
MC G - 0.11$^{(f)}$  & 2.8  & 4.3  & 1.6  & -- & 15 \\
MC G - 0.02$^{(f)}$  & 2.5   &  3.3  & 1.3 & -- & 15 \\
Lemmon$^{(g)}$ & $< 0.7$  & --  & $ > 3$  & -- & 16 \\
LoveJoy$^{(g)}$ & $< 0.6$  & --  & $> 5$ & -- & 16  \\
Hale-Bopp$^{(g)}$ & 0.32  & $>2$   & $>$ 6.25 & -- & 17 \\

\hline
\end{tabular}
\tablefoot{The table show the results from similar studies of : $^{(a)}$ The Orion Kleinmann-Low nebula, a high-mass star-forming region $^{(b)}$ the Large Molecule Heimat (LMH) hot core source toward Sgr B2(N), $^{(c)}$high-mass star-forming region, $^{(d)}$ intermediate-mass protostar , $^{(e)}$ low-mass Class 0 protostars, $^{(f)}$ molecular clouds in the Sgr A complex (MC~G~-~0.02~-~0.07 ("the 50~km~s$^{-1}$ cloud") and MC~G~-~0.11~-~0.08 ("the 20~km~s$^{-1}$ cloud")) and in the Sgr~B2 complex (MC~G~+~0.693~-~0.03) and $^{(g)}$ comets.}
\tablebib{(1)~this study; (2)\citet{Urquhart2014}; (3)~\citet{Favre2011,Brouillet2015}; (4)~\citet{Crockett2014};
(5)~\citet{vanDishoeck2011}; (6)~\citet{Hollis2001}; (7)~\citet{Goldsmith1990}; (8)~\citet{Beltran2009}; (9)~\citet{Fuente2014};(10)~\citet{Coutens2015}; (11)~\citet{Jorgensen2009}; (12)~Jørgensen et al. (2012, in prep.);
(13)~\citet{Schoier2002}; (14)~\citet{Taquet2015}; (15)~\citet{Requena2008}; (16)~\citet{Biver2014}; (17)~\citet{Bockel2000,Crovisier2004a,Crovisier2004b}.
}
\end{table*}

\section{Discussion}\label{sec:disc}
\EG{} has previously been detected in several other sources (Sgr B2(N), NGC~7129~FIRS2, NGC~1333~IRAS~2A and Orion-KL), with additional marginal/tentative detections in W51 e1/e2 and IRAS~16293--2422 \citep{Hollis2002,Fuente2014,Maury2014,Coutens2015,Kalenskii2010,Jorgensen2012}. In this paper we have identified \EG{} and \MF{} in two high-mass protostellar sources, W51/2 and G34.3+0.2, and in addition been able to give an upper limit for the column density of \GLY{} in the same sources. The results from this study are roughly similar to the estimates ratios seen in Orion~KL \citep{Favre2011,Brouillet2015} for the HCOOCH$_3$/CH$_2$OHCHO and HCOOCH$_3$/(CH$_2$OH)$_2$ ratio (table~\ref{tcompare}) while the upper limits for (CH$_2$OH)$_2$/CH$_2$OHCHO are similar to the results for NGC~1333~IRAS~2A \citep{Coutens2015} as well as the upper limits results for comets \citep{Biver2014,Bockel2000,Crovisier2004a,Crovisier2004b}.

\begin{figure}
\centering
\includegraphics[width = 8.5cm]{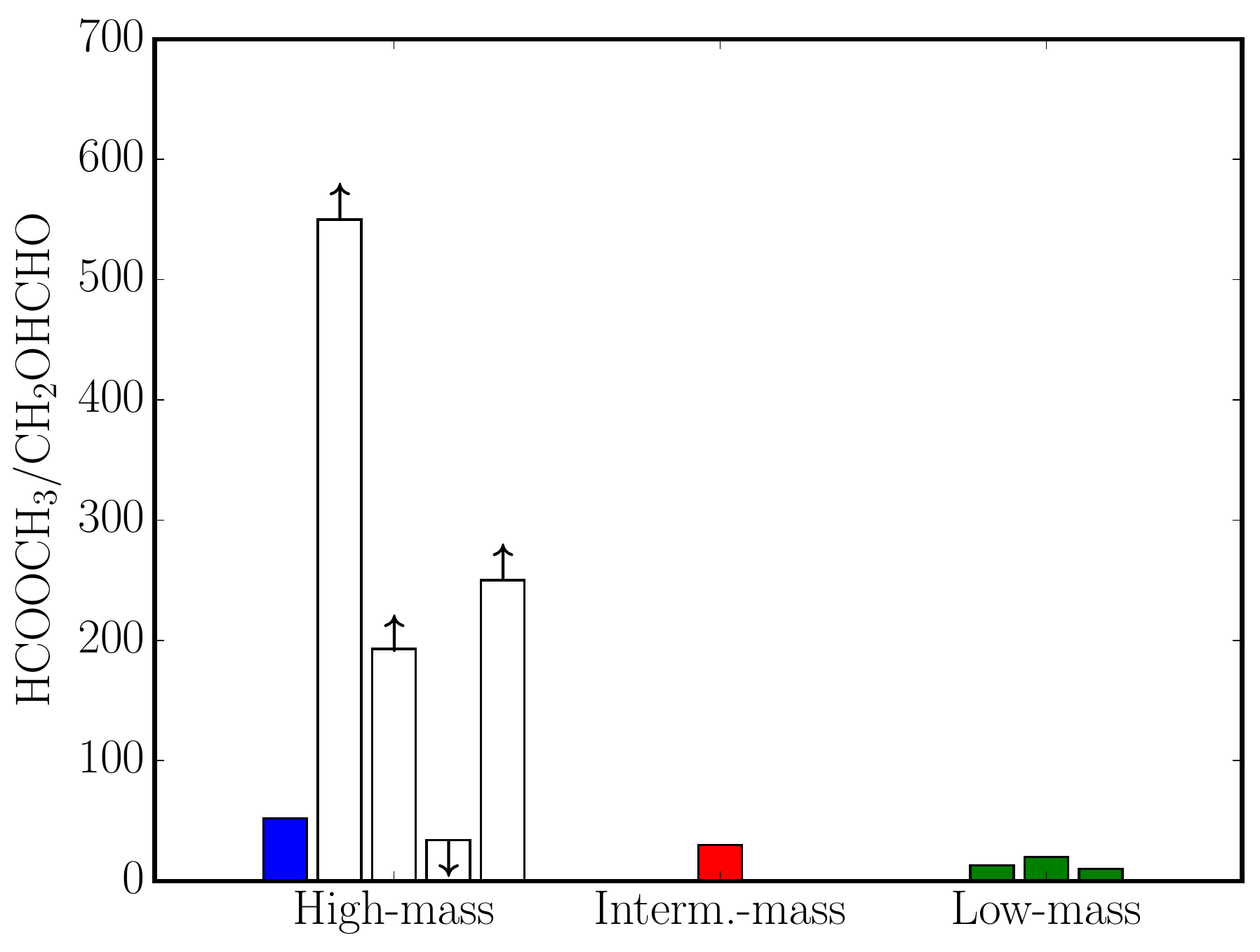}
\includegraphics[width = 8.5cm]{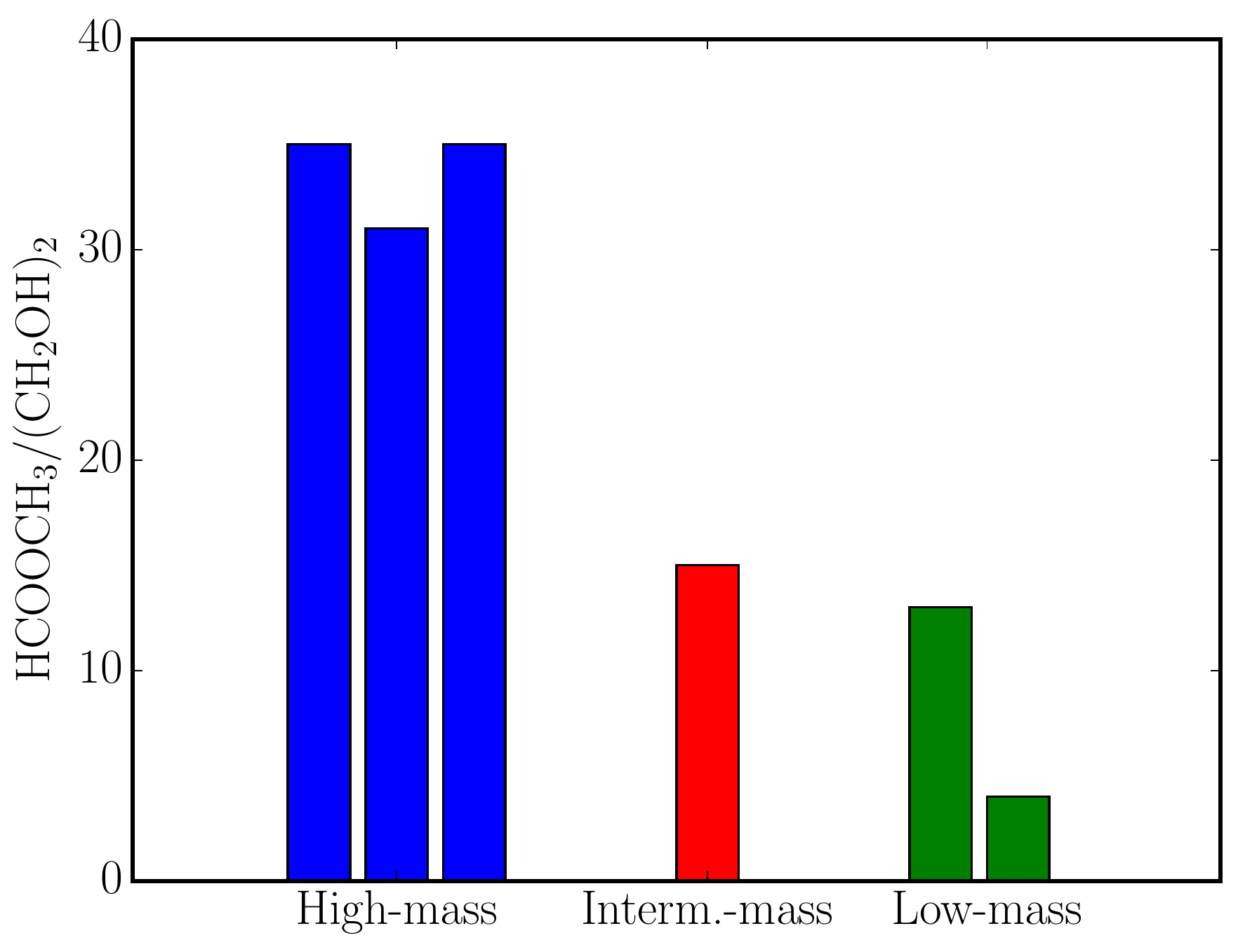}
\caption{\label{Bar_plot}Schematic bar-plot of HCOOCH$_3$/CH$_2$OHCHO (top) and HCOOCH$_3$/(CH$_2$OH)$_2$ (bottom) against $L_\mathrm{bol}$. The sources from Table~\ref{tcompare} have been plotted in descending order of luminosity from left to right on the x-axis, but not to scale, and the sources have been grouped into high-mass sources (blue), intermediate-mass protostar (red) and low-mass protostars (green). The white bars for four of the high-mass sources are upper/lower levels, which is indicated by the direction of the arrow. The luminosity for the top plot spans from Sgr~B2(N) with $L_\mathrm{bol} = 10^7 L_{\sun}$ to IRAS NGC 1333 4A with $L_\mathrm{bol} = 7.7 L_{\sun}$, while it spans from W51/e2 with $L_\mathrm{bol} = 4.7 \times 10^6 L_{\sun}$ to IRAS NGC 1333 4A with $L_\mathrm{bol} = 20 L_{\sun}$ in the bottom plot.}
\end{figure} 

When investigating the conditions leading to differences in observed abundance ratios, it is important to take both the formation process as well as the destruction processes of the molecules into consideration. \cite{Garrod2008} combine a gas-grain chemical network with a physical model to test if the chemistry can reproduce the observed abundances of previously detected organic molecules in physical conditions characteristic to star-forming regions. The physical model used in \cite{Garrod2008} is based on \cite{Viti2004} and consist of an isothermal collapse followed by a warm-up phase with temperatures from 10~K to 200~K, assuming absolute timescales for the warm-up phases are $1 \times 10^{6}$, $2 \times 10^{5}$ and $5 \times 10^{4}$~yr representing low-, intermediate- and high-mass star-formation respectively. However \cite{Aikawa2008} argue that the relation should be reversed, as the warm-up timescale should be relative and should depend on the ratio of the size of the warm region to the infall speed instead of the overall speed of star formation. Either way, \cite{Garrod2008} and \cite{Aikawa2008} agree that the timescales, whether absolute or relative, at the different temperature-ranges are important for the chemistry. 

For all the sources in Table~\ref{tcompare}, \MF{} is more abundant than \GLY{}, from a factor of >~550 in W51/e2 to a factor of  >~2 in comet Hale-Bopp. According to \cite{Garrod2008} \MF{} and \GLY{} have similar formation pathways, which are based on the addition of HCO and CH$_3$O or CH$_2$OH  at 30--40~K. \cite{Garrod2008} find that the production rates of CH$_3$O and CH$_2$OH are the same, which leads to similar abundances for \MF{} and \GLY{}. Intuitively this makes sense as the two molecules are isomers but it contradicts observations as \MF{} is observed to be much more abundant than \GLY{} in all the sources reported so far. As suggested by \cite{Garrod2008}, this discrepancy could be due to the differences in the CH$_3$O/CH$_2$OH branching ratio. According to \cite{Oberg2009}, \MF{} forms at a lower temperature than \GLY{}, which can also have a impact on the resulting ratio of the two species. Another explanation to the large HCOOCH$_3$/CH$_2$OHCHO ratio is the assumptions regarding thermal evaporation for these two species.  \cite{Garrod2008} suggest that \GLY{} remains on the grains until is co-desorbs with water (at $\sim$ 110 K) while \MF{} evaporates at 70--80~K. This leaves \GLY{} to be destroyed by OH radicals at higher temperatures on the grains prior to evaporation. As \cite{Oberg2009} conclude, the ratio of HCOOCH$_3$/CH$_2$OHCHO does not depend greatly on the initial ice composition, and so the observed variations will most likely be linked to the different temperature conditions of the different sources. 

In Table~\ref{tcompare} the sources are listed in order with decreasing HCOOCH$_3$/CH$_2$OHCHO ratio, which in turn also roughly correspond to a descending order of luminosity, which is also listed in the table. The top plot in Figure~\ref{Bar_plot} shows a schematic bar-plot of HCOOCH$_3$/CH$_2$OHCHO against $L_\mathrm{bol}$. On the x-axis the sources in Table~\ref{tcompare} have been plotted in descending order from left to right, but not to scale. As illustrated by Figure~\ref{Bar_plot} (top) a rough correlation between HCOOCH$_3$/CH$_2$OHCHO and $L_\mathrm{bol}$ exist: more luminous sources show a high HCOOCH$_3$/CH$_2$OHCHO ratio, while low luminosity sources show a low HCOOCH$_3$/CH$_2$OHCHO ratio with intermediate values in between. The apparent correlation between HCOOCH$_3$/CH$_2$OHCHO and source luminosity supports the hypothesis that the HCOOCH$_3$/CH$_2$OHCHO ratio depends mainly on the temperature, which in turns depends on the luminosity of the source. Even if the chemistry and the temperature profile in all the sources were quite similar, the differences in timescales at the different temperature-ranges (low, intermediate to high temperature) are not. A higher HCOOCH$_3$/CH$_2$OHCHO ratio in more luminous sources, could simply be a consequence of \textit{i)} more luminous sources have experienced a longer timescale at temperatures which are either more favorable to the formation of \MF{} and/or to the destruction of \GLY{} or \textit{ii)} less luminous sources experience a shorter timescale at high temperature, which would conserve a greater fraction of \GLY{} than their more luminous counterparts. One should of course keep in mind that the sources discussed here have temperature profiles which covers a range of temperatures falling of with radius and that hot core regions are permeated by shocks which also produce a range of temperatures. For the Galactic Centre molecular clouds from the study by \cite{Requena2008} no luminosities are listed: these are warm, low density clouds with no sign of star-formation. However, judging by the HCOOCH$_3$/CH$_2$OHCHO and HCOOCH$_3$/(CH$_2$OH)$_2$ ratios, they appear to be closer to the low-mass protostars than to the hot cores and high-mass star-forming regions. 

In the bottom plot in Figure~\ref{Bar_plot}, the HCOOCH$_3$/(CH$_2$OH)$_2$ abundance ratio also shows a decrease with $L_\mathrm{bol}$ similar to that of HCOOCH$_3$/CH$_2$OHCHO. If the same temperature timescale argument is applied to this correlation, one would expect an opposite trend, as \EG{} is formed a high temperatures \citep{Garrod2008}. A possible explanation for this seemingly contradiction is that less luminous sources might experience high temperatures at timescales, that are just long enough for \EG{} to form, but not long enough for \EG{} to be destroyed again. Another possible explanation is that the initial ice composition affects the abundance ratio. While the HCOOCH$_3$/CH$_2$OHCHO ratio is independent of initial ice composition, then the formation of \EG{} is strongly correlated to the CH$_3$OH:CO composition of the ice \citep{Oberg2009}. \cite{Oberg2009} show that pure CH$_3$OH ice strongly enhances the \EG{} abundance as compared to ice mixes containing CO. This contradicts the predictions by \cite{Garrod2008} where the \EG{} abundance actually drops 1-2 magnitudes when the initial CH$_3$OH ice-composition in their model is reduced by a factor of 10. However, to date there is no observational evidence that the ice contents on the grains vary statistically significant from high-mass protostars to low-mass protostars \citep{Oberg2011}. 

Until recently a high (CH$_2$OH)$_2$/CH$_2$OHCHO ratio was speculated to be specific property of comets \citep{Biver2014}. But as \citet{Coutens2015} has reported a values of $\sim$ 5 for a low-mass protostar and we, together with \cite{Brouillet2015}, report lower limits of > 6--16 for high-mass sources, then a (CH$_2$OH)$_2$/CH$_2$OHCHO ratio larger than 3 can be expected to be observed in other sources as well.

\section{Conclusions}
\label{sec:conc}

In summary, we have tentatively detected \EG{} in G34.3+0.2 for the first time and our tentative detection in W51/e2 confirms the previous marginally detection by \citet{Kalenskii2010}. In addition, we derive upper limits for the column density of \GLY{} emission in both sources. From these data we determine the relative abundances of \EG{}, x and \GLY{}. The relative abundances of these species are compared to measurements from literature covering a wide range of source environments and luminosities. The data show what appears to be a correlation between source luminosity and HCOOCH$_3$/(CH$_2$OH)$_2$ as well as HCOOCH$_3$/CH$_2$OHCHO. This apparent correlation is proposed to be a consequence of the relative timescales each source spend at different temperatures-ranges in their evolution. Using the upper limit for the column density of \GLY{} gives a lower limit for (CH$_2$OH)$_2$/CH$_2$OHCHO of $ > 16$ and $ > 6$ for W51/2 and G34.3+0.2 respectively. These results, together with an upper limit of $ > 12$ for Orion-KL \citep{Brouillet2015} and (CH$_2$OH)$_2$/CH$_2$OHCHO $ = 5$ for IRAS NGC 1333 2A \citep{Coutens2015}, shows that one can expect to find high \EG{}/\GLY{} abundance ratios in multiple types of environments. 

Additional systematic surveys of these and other relevant molecules in additional sources as well as more model and laboratory work is needed in order to fully constrain the formation pathway of complex molecules. In particular, the Atacama Large Millimeter/submillimeter Array (ALMA) shows great potential for successfully revealing the formation processes with its high sensitivity and resolution making it possible to map the relative spatial distributions of these sources.

\begin{acknowledgements}
The research of JML and JKJ was supported by a Junior Group Leader Fellowship from the Lundbeck Foundation to JKJ -- as well as Centre for Star and Planet Formation funded by the Danish National Research Foundation. CF acknowledges support from the National Science Foundation under grant 1008800. 
\end{acknowledgements}



\bibliographystyle{aa}
\bibliography{new.bib}

\begin{thebibliography}{23}
\expandafter\ifx\csname natexlab\endcsname\relax\def\natexlab#1{#1}\fi

\bibitem[{{Beuther} {et~al.}(2005){Beuther}, {Zhang}, {Greenhill}, {Reid},
  {Wilner}, {Keto}, {Shinnaga}, {Ho}, {Moran}, {Liu}, \&
  {Chang}}]{Beuther:2005}
{Beuther}, H., {Zhang}, Q., {Greenhill}, L.~J., {et~al.} 2005, \apj, 632, 355

\bibitem[{{Blake} {et~al.}(1987){Blake}, {Sutton}, {Masson}, \&
  {Phillips}}]{Blake:1987}
{Blake}, G.~A., {Sutton}, E.~C., {Masson}, C.~R., \& {Phillips}, T.~G. 1987,
  \apj, 315, 621

\bibitem[{{Combes} {et~al.}(1996){Combes}, {Q-Rieu}, \&
  {Wlodarczak}}]{Combes:1996}
{Combes}, F., {Q-Rieu}, N., \& {Wlodarczak}, G. 1996, \aap, 308, 618

\bibitem[{{Comito} {et~al.}(2005){Comito}, {Schilke}, {Phillips}, {Lis},
  {Motte}, \& {Mehringer}}]{Comito:2005}
{Comito}, C., {Schilke}, P., {Phillips}, T.~G., {et~al.} 2005, \apjs, 156, 127

\bibitem[{{de Vicente} {et~al.}(2002){de Vicente}, {Mart{\'{\i}}n-Pintado},
  {Neri}, \& {Rodr{\'{\i}}guez-Franco}}]{de-Vicente:2002}
{de Vicente}, P., {Mart{\'{\i}}n-Pintado}, J., {Neri}, R., \&
  {Rodr{\'{\i}}guez-Franco}, A. 2002, \apjl, 574, L163

\bibitem[{{Draine} \& {Lee}(1984)}]{Draine:1984}
{Draine}, B.~T. \& {Lee}, H.~M. 1984, \apj, 285, 89

\bibitem[{{Eisner} {et~al.}(2008){Eisner}, {Plambeck}, {Carpenter}, {Corder},
  {Qi}, \& {Wilner}}]{Eisner:2008}
{Eisner}, J.~A., {Plambeck}, R.~L., {Carpenter}, J.~M., {et~al.} 2008, \apj,
  683, 304

\bibitem[{{Forster} {et~al.}(1978){Forster}, {Welch}, {Wright}, \&
  {Baudry}}]{Forster:1978}
{Forster}, J.~R., {Welch}, W.~J., {Wright}, M.~C.~H., \& {Baudry}, A. 1978,
  \apj, 221, 137

\bibitem[{{Friedel} \& {Snyder}(2008)}]{Friedel:2008}
{Friedel}, D.~N. \& {Snyder}, L.~E. 2008, \apj, 672, 962

\bibitem[{{Goddi} {et~al.}(2009){Goddi}, {Greenhill}, {Chandler}, {Humphreys},
  {Matthews}, \& {Gray}}]{Goddi:2009}
{Goddi}, C., {Greenhill}, L.~J., {Chandler}, C.~J., {et~al.} 2009, \apj, 698,
  1165

\bibitem[{{G{\'o}mez} {et~al.}(2005){G{\'o}mez}, {Rodr{\'{\i}}guez}, {Loinard},
  {Lizano}, {Poveda}, \& {Allen}}]{Gomez:2005}
{G{\'o}mez}, L., {Rodr{\'{\i}}guez}, L.~F., {Loinard}, L., {et~al.} 2005, \apj,
  635, 1166

\bibitem[{{Gu{\'e}lin} {et~al.}(2008){Gu{\'e}lin}, {Brouillet}, {Cernicharo},
  {Combes}, \& {Wooten}}]{Guelin:2008}
{Gu{\'e}lin}, M., {Brouillet}, N., {Cernicharo}, J., {Combes}, F., \& {Wooten},
  A. 2008, \apss, 313, 45

\bibitem[{{Ilyushin} {et~al.}(2009){Ilyushin}, {Kryvda}, \&
  {Alekseev}}]{Ilyushin:2009}
{Ilyushin}, V., {Kryvda}, A., \& {Alekseev}, E. 2009, Journal of Molecular
  Spectroscopy, 255, 32

\bibitem[{{Kobayashi} {et~al.}(2007){Kobayashi}, {Ogata}, {Tsunekawa}, \&
  {Takano}}]{Kobayashi:2007}
{Kobayashi}, K., {Ogata}, K., {Tsunekawa}, S., \& {Takano}, S. 2007, \apjl,
  657, L17

\bibitem[{{Mathis} {et~al.}(1977){Mathis}, {Rumpl}, \&
  {Nordsieck}}]{Mathis:1977}
{Mathis}, J.~S., {Rumpl}, W., \& {Nordsieck}, K.~H. 1977, \apj, 217, 425

\bibitem[{{Menten} \& {Reid}(1995)}]{Menten:1995}
{Menten}, K.~M. \& {Reid}, M.~J. 1995, \apjl, 445, L157

\bibitem[{{Menten} {et~al.}(2007){Menten}, {Reid}, {Forbrich}, \&
  {Brunthaler}}]{Menten:2007}
{Menten}, K.~M., {Reid}, M.~J., {Forbrich}, J., \& {Brunthaler}, A. 2007, \aap,
  474, 515

\bibitem[{{Pickett} {et~al.}(1998){Pickett}, {Poynter}, {Cohen}, {Delitsky},
  {Pearson}, \& {Muller}}]{Pickett:1998}
{Pickett}, H.~M., {Poynter}, I.~R.~L., {Cohen}, E.~A., {et~al.} 1998, Journal
  of Quantitative Spectroscopy and Radiative Transfer, 60, 883

\bibitem[{{Plambeck} {et~al.}(1995){Plambeck}, {Wright}, {Mundy}, \&
  {Looney}}]{Plambeck:1995}
{Plambeck}, R.~L., {Wright}, M.~C.~H., {Mundy}, L.~G., \& {Looney}, L.~W. 1995,
  \apjl, 455, L189+

\bibitem[{{Rodr{\'{\i}}guez} {et~al.}(2005){Rodr{\'{\i}}guez}, {Poveda},
  {Lizano}, \& {Allen}}]{Rodriguez:2005}
{Rodr{\'{\i}}guez}, L.~F., {Poveda}, A., {Lizano}, S., \& {Allen}, C. 2005,
  \apjl, 627, L65

\bibitem[{{Turner}(1991)}]{Turner:1991}
{Turner}, B.~E. 1991, \apjs, 76, 617

\bibitem[{{Wilson} {et~al.}(2000){Wilson}, {Gaume}, {Gensheimer}, \&
  {Johnston}}]{Wilson:2000}
{Wilson}, T.~L., {Gaume}, R.~A., {Gensheimer}, P., \& {Johnston}, K.~J. 2000,
  \apj, 538, 665

\bibitem[{{Wynn-Williams} {et~al.}(1984){Wynn-Williams}, {Genzel}, {Becklin},
  \& {Downes}}]{Wynn-Williams:1984}
{Wynn-Williams}, C.~G., {Genzel}, R., {Becklin}, E.~E., \& {Downes}, D. 1984,
  \apj, 281, 172

\end{thebibliography}


\begin{thebibliography}{50}
\expandafter\ifx\csname natexlab\endcsname\relax\def\natexlab#1{#1}\fi

\bibitem[{{Aikawa} {et~al.}(2008){Aikawa}, {Wakelam}, {Garrod}, \&
  {Herbst}}]{Aikawa2008}
{Aikawa}, Y., {Wakelam}, V., {Garrod}, R.~T., \& {Herbst}, E. 2008, \apj, 674,
  984

\bibitem[{{Beltr{\'a}n} {et~al.}(2009){Beltr{\'a}n}, {Codella}, {Viti}, {Neri},
  \& {Cesaroni}}]{Beltran2009}
{Beltr{\'a}n}, M.~T., {Codella}, C., {Viti}, S., {Neri}, R., \& {Cesaroni}, R.
  2009, \apjl, 690, L93

\bibitem[{{Bisschop} {et~al.}(2007){Bisschop}, {J{\o}rgensen}, {van Dishoeck},
  \& {de Wachter}}]{Bisschop2007}
{Bisschop}, S.~E., {J{\o}rgensen}, J.~K., {van Dishoeck}, E.~F., \& {de
  Wachter}, E.~B.~M. 2007, \aap, 465, 913

\bibitem[{{Biver} {et~al.}(2014){Biver}, {Bockel{\'e}e-Morvan}, {Debout},
  {Crovisier}, {Boissier}, {Lis}, {Dello Russo}, {Moreno}, {Colom}, {Paubert},
  {Vervack}, \& {Weaver}}]{Biver2014}
{Biver}, N., {Bockel{\'e}e-Morvan}, D., {Debout}, V., {et~al.} 2014, \aap, 566,
  L5

\bibitem[{{Blake} {et~al.}(1987){Blake}, {Sutton}, {Masson}, \&
  {Phillips}}]{Blake:1987}
{Blake}, G.~A., {Sutton}, E.~C., {Masson}, C.~R., \& {Phillips}, T.~G. 1987,
  \apj, 315, 621

\bibitem[{{Bockel{\'e}e-Morvan} {et~al.}(2000){Bockel{\'e}e-Morvan}, {Lis},
  {Wink}, {Despois}, {Crovisier}, {Bachiller}, {Benford}, {Biver}, {Colom},
  {Davies}, {G{\'e}rard}, {Germain}, {Houde}, {Mehringer}, {Moreno}, {Paubert},
  {Phillips}, \& {Rauer}}]{Bockel2000}
{Bockel{\'e}e-Morvan}, D., {Lis}, D.~C., {Wink}, J.~E., {et~al.} 2000, \aap,
  353, 1101

\bibitem[{{Brouillet} {et~al.}(2015){Brouillet}, {Despois}, {Lu}, {Baudry},
  {Cernicharo}, {Bockel{\'e}e-Morvan}, {Crovisier}, \& {Biver}}]{Brouillet2015}
{Brouillet}, N., {Despois}, D., {Lu}, X.-H., {et~al.} 2015, \aap, 576, A129

\bibitem[{{Carroll} {et~al.}(2010){Carroll}, {Drouin}, \& {Widicus
  Weaver}}]{Carroll2010}
{Carroll}, P.~B., {Drouin}, B.~J., \& {Widicus Weaver}, S.~L. 2010, \apj, 723,
  845

\bibitem[{{Charnley} {et~al.}(1992){Charnley}, {Tielens}, \&
  {Millar}}]{Charnley1992}
{Charnley}, S.~B., {Tielens}, A.~G.~G.~M., \& {Millar}, T.~J. 1992, \apjl, 399,
  L71

\bibitem[{{Christen} {et~al.}(1995){Christen}, {Coudert}, {Suenram}, \&
  {Lovas}}]{Christen:1995}
{Christen}, D., {Coudert}, L.~H., {Suenram}, R.~D., \& {Lovas}, F.~J. 1995,
  Journal of Molecular Spectroscopy, 172, 57

\bibitem[{{Christen} \& {M{\"u}ller}(2003)}]{Christen:2003}
{Christen}, D. \& {M{\"u}ller}, H.~S.~P. 2003, Physical Chemistry Chemical
  Physics (Incorporating Faraday Transactions), 5, 3600

\bibitem[{{Coutens} {et~al.}(2015){Coutens}, {Persson}, {J{\o}rgensen},
  {Wampfler}, \& {Lykke}}]{Coutens2015}
{Coutens}, A., {Persson}, M.~V., {J{\o}rgensen}, J.~K., {Wampfler}, S.~F., \&
  {Lykke}, J.~M. 2015, \aap, 576, A5

\bibitem[{{Crockett} {et~al.}(2014){Crockett}, {Bergin}, {Neill}, {Black},
  {Blake}, \& {Kleshcheva}}]{Crockett2014}
{Crockett}, N.~R., {Bergin}, E.~A., {Neill}, J.~L., {et~al.} 2014, \apj, 781,
  114

\bibitem[{{Crovisier} {et~al.}(2004{\natexlab{a}}){Crovisier},
  {Bockel{\'e}e-Morvan}, {Biver}, {Colom}, {Despois}, \&
  {Lis}}]{Crovisier2004a}
{Crovisier}, J., {Bockel{\'e}e-Morvan}, D., {Biver}, N., {et~al.}
  2004{\natexlab{a}}, \aap, 418, L35

\bibitem[{{Crovisier} {et~al.}(2004{\natexlab{b}}){Crovisier},
  {Bockel{\'e}e-Morvan}, {Colom}, {Biver}, {Despois}, {Lis}, \& {Team for
  target-of-opportunity radio observations of comets}}]{Crovisier2004b}
{Crovisier}, J., {Bockel{\'e}e-Morvan}, D., {Colom}, P., {et~al.}
  2004{\natexlab{b}}, \aap, 418, 1141

\bibitem[{{Demyk} {et~al.}(2008){Demyk}, {Wlodarczak}, \&
  {Carvajal}}]{Demyk:2008}
{Demyk}, K., {Wlodarczak}, G., \& {Carvajal}, M. 2008, \aap, 489, 589

\bibitem[{{Favre} {et~al.}(2014){Favre}, {Carvajal}, {Field}, {J{\o}rgensen},
  {Bisschop}, {Brouillet}, {Despois}, {Baudry}, {Kleiner}, {Bergin},
  {Crockett}, {Neill}, {Margul{\`e}s}, {Huet}, \& {Demaison}}]{favre:2014a}
{Favre}, C., {Carvajal}, M., {Field}, D., {et~al.} 2014, \apjs, 215, 25

\bibitem[{{Favre} {et~al.}(2011){Favre}, {Despois}, {Brouillet}, {Baudry},
  {Combes}, {Gu{\'e}lin}, {Wootten}, \& {Wlodarczak}}]{Favre2011}
{Favre}, C., {Despois}, D., {Brouillet}, N., {et~al.} 2011, \aap, 532, A32

\bibitem[{{Fuente} {et~al.}(2014){Fuente}, {Cernicharo}, {Caselli}, {McCoey},
  {Johnstone}, {Fich}, {van Kempen}, {Palau}, {Y{\i}ld{\i}z}, {Tercero}, \&
  {L{\'o}pez}}]{Fuente2014}
{Fuente}, A., {Cernicharo}, J., {Caselli}, P., {et~al.} 2014, \aap, 568, A65

\bibitem[{{Garrod} {et~al.}(2008){Garrod}, {Weaver}, \& {Herbst}}]{Garrod2008}
{Garrod}, R.~T., {Weaver}, S.~L.~W., \& {Herbst}, E. 2008, \apj, 682, 283

\bibitem[{{Geppert} {et~al.}(2006){Geppert}, {Hamberg}, {Thomas},
  {{\"O}sterdahl}, {Hellberg}, {Zhaunerchyk}, {Ehlerding}, {Millar}, {Roberts},
  {Semaniak}, {Ugglas}, {K{\"a}llberg}, {Simonsson}, {Kaminska}, \&
  {Larsson}}]{Geppert2006}
{Geppert}, W.~D., {Hamberg}, M., {Thomas}, R.~D., {et~al.} 2006, Faraday
  Discussions, 133, 177

\bibitem[{{Goldsmith} \& {Langer}(1999)}]{Goldsmith1999}
{Goldsmith}, P.~F. \& {Langer}, W.~D. 1999, \apj, 517, 209

\bibitem[{{Goldsmith} {et~al.}(1990){Goldsmith}, {Lis}, {Hills}, \&
  {Lasenby}}]{Goldsmith1990}
{Goldsmith}, P.~F., {Lis}, D.~C., {Hills}, R., \& {Lasenby}, J. 1990, \apj,
  350, 186

\bibitem[{{Halfen} {et~al.}(2006){Halfen}, {Apponi}, {Woolf}, {Polt}, \&
  {Ziurys}}]{Halfen2006}
{Halfen}, D.~T., {Apponi}, A.~J., {Woolf}, N., {Polt}, R., \& {Ziurys}, L.~M.
  2006, \apj, 639, 237

\bibitem[{{Herbst} \& {van Dishoeck}(2009)}]{Herbst2009}
{Herbst}, E. \& {van Dishoeck}, E.~F. 2009, \araa, 47, 427

\bibitem[{{Hollis} {et~al.}(2004){Hollis}, {Jewell}, {Lovas}, \&
  {Remijan}}]{Hollis2004}
{Hollis}, J.~M., {Jewell}, P.~R., {Lovas}, F.~J., \& {Remijan}, A. 2004, \apjl,
  613, L45

\bibitem[{{Hollis} {et~al.}(2000){Hollis}, {Lovas}, \& {Jewell}}]{Hollis2000}
{Hollis}, J.~M., {Lovas}, F.~J., \& {Jewell}, P.~R. 2000, \apjl, 540, L107

\bibitem[{{Hollis} {et~al.}(2002){Hollis}, {Lovas}, {Jewell}, \&
  {Coudert}}]{Hollis2002}
{Hollis}, J.~M., {Lovas}, F.~J., {Jewell}, P.~R., \& {Coudert}, L.~H. 2002,
  \apjl, 571, L59

\bibitem[{{Hollis} {et~al.}(2001){Hollis}, {Vogel}, {Snyder}, {Jewell}, \&
  {Lovas}}]{Hollis2001}
{Hollis}, J.~M., {Vogel}, S.~N., {Snyder}, L.~E., {Jewell}, P.~R., \& {Lovas},
  F.~J. 2001, \apjl, 554, L81

\bibitem[{{Ilyushin} {et~al.}(2009){Ilyushin}, {Kryvda}, \&
  {Alekseev}}]{Ilyushin:2009}
{Ilyushin}, V., {Kryvda}, A., \& {Alekseev}, E. 2009, Journal of Molecular
  Spectroscopy, 255, 32

\bibitem[{{J{\o}rgensen} {et~al.}(2012){J{\o}rgensen}, {Favre}, {Bisschop},
  {Bourke}, {van Dishoeck}, \& {Schmalzl}}]{Jorgensen2012}
{J{\o}rgensen}, J.~K., {Favre}, C., {Bisschop}, S.~E., {et~al.} 2012, \apjl,
  757, L4

\bibitem[{{J{\o}rgensen} {et~al.}(2009){J{\o}rgensen}, {van Dishoeck},
  {Visser}, {Bourke}, {Wilner}, {Lommen}, {Hogerheijde}, \&
  {Myers}}]{Jorgensen2009}
{J{\o}rgensen}, J.~K., {van Dishoeck}, E.~F., {Visser}, R., {et~al.} 2009,
  \aap, 507, 861

\bibitem[{{Kalenskii} \& {Johansson}(2010)}]{Kalenskii2010}
{Kalenskii}, S.~V. \& {Johansson}, L.~E.~B. 2010, Astronomy Reports, 54, 1084

\bibitem[{{Kurtz} {et~al.}(2000){Kurtz}, {Cesaroni}, {Churchwell}, {Hofner}, \&
  {Walmsley}}]{Kurtz2000}
{Kurtz}, S., {Cesaroni}, R., {Churchwell}, E., {Hofner}, P., \& {Walmsley},
  C.~M. 2000, Protostars and Planets IV, 299

\bibitem[{{Maury} {et~al.}(2014){Maury}, {Belloche}, {Andr{\'e}}, {Maret},
  {Gueth}, {Codella}, {Cabrit}, {Testi}, \& {Bontemps}}]{Maury2014}
{Maury}, A.~J., {Belloche}, A., {Andr{\'e}}, P., {et~al.} 2014, \aap, 563, L2

\bibitem[{{Millar} {et~al.}(1991){Millar}, {Herbst}, \&
  {Charnley}}]{Millar1991}
{Millar}, T.~J., {Herbst}, E., \& {Charnley}, S.~B. 1991, \apj, 369, 147

\bibitem[{{M{\"u}ller} {et~al.}(2005){M{\"u}ller}, {Schl{\"o}der}, {Stutzki},
  \& {Winnewisser}}]{Muller2005}
{M{\"u}ller}, H.~S.~P., {Schl{\"o}der}, F., {Stutzki}, J., \& {Winnewisser}, G.
  2005, Journal of Molecular Structure, 742, 215

\bibitem[{{M{\"u}ller} {et~al.}(2001){M{\"u}ller}, {Thorwirth}, {Roth}, \&
  {Winnewisser}}]{Muller2001}
{M{\"u}ller}, H.~S.~P., {Thorwirth}, S., {Roth}, D.~A., \& {Winnewisser}, G.
  2001, \aap, 370, L49

\bibitem[{{{\"O}berg} {et~al.}(2011){{\"O}berg}, {Boogert}, {Pontoppidan}, {van
  den Broek}, {van Dishoeck}, {Bottinelli}, {Blake}, \& {Evans}}]{Oberg2011}
{{\"O}berg}, K.~I., {Boogert}, A.~C.~A., {Pontoppidan}, K.~M., {et~al.} 2011,
  \apj, 740, 109

\bibitem[{{{\"O}berg} {et~al.}(2009){{\"O}berg}, {Garrod}, {van Dishoeck}, \&
  {Linnartz}}]{Oberg2009}
{{\"O}berg}, K.~I., {Garrod}, R.~T., {van Dishoeck}, E.~F., \& {Linnartz}, H.
  2009, \aap, 504, 891

\bibitem[{{Pickett} {et~al.}(1998){Pickett}, {Poynter}, {Cohen}, {Delitsky},
  {Pearson}, \& {Muller}}]{Pickett:1998}
{Pickett}, H.~M., {Poynter}, I.~R.~L., {Cohen}, E.~A., {et~al.} 1998, Journal
  of Quantitative Spectroscopy and Radiative Transfer, 60, 883

\bibitem[{{Remijan} {et~al.}(2003){Remijan}, {Snyder}, {Friedel}, {Liu}, \&
  {Shah}}]{Remijan2003}
{Remijan}, A., {Snyder}, L.~E., {Friedel}, D.~N., {Liu}, S.-Y., \& {Shah},
  R.~Y. 2003, \apj, 590, 314

\bibitem[{{Requena-Torres} {et~al.}(2008){Requena-Torres},
  {Mart{\'{\i}}n-Pintado}, {Mart{\'{\i}}n}, \& {Morris}}]{Requena2008}
{Requena-Torres}, M.~A., {Mart{\'{\i}}n-Pintado}, J., {Mart{\'{\i}}n}, S., \&
  {Morris}, M.~R. 2008, \apj, 672, 352

\bibitem[{{Sato} {et~al.}(2010){Sato}, {Reid}, {Brunthaler}, \&
  {Menten}}]{Sato2010}
{Sato}, M., {Reid}, M.~J., {Brunthaler}, A., \& {Menten}, K.~M. 2010, \apj,
  720, 1055

\bibitem[{{Sch{\"o}ier} {et~al.}(2002){Sch{\"o}ier}, {J{\o}rgensen}, {van
  Dishoeck}, \& {Blake}}]{Schoier2002}
{Sch{\"o}ier}, F.~L., {J{\o}rgensen}, J.~K., {van Dishoeck}, E.~F., \& {Blake},
  G.~A. 2002, \aap, 390, 1001

\bibitem[{{Taquet} {et~al.}(2015){Taquet}, {L{\'o}pez-Sepulcre}, {Ceccarelli},
  {Neri}, {Kahane}, \& {Charnley}}]{Taquet2015}
{Taquet}, V., {L{\'o}pez-Sepulcre}, A., {Ceccarelli}, C., {et~al.} 2015, \apj,
  804, 81

\bibitem[{{Urquhart} {et~al.}(2014){Urquhart}, {Figura}, {Moore}, {Hoare},
  {Lumsden}, {Mottram}, {Thompson}, \& {Oudmaijer}}]{Urquhart2014}
{Urquhart}, J.~S., {Figura}, C.~C., {Moore}, T.~J.~T., {et~al.} 2014, \mnras,
  437, 1791

\bibitem[{{van Dishoeck} {et~al.}(2011){van Dishoeck}, {Kristensen}, {Benz},
  {Bergin}, {Caselli}, {Cernicharo}, {Herpin}, {Hogerheijde}, {Johnstone},
  {Liseau}, {Nisini}, {Shipman}, {Tafalla}, {van der Tak}, {Wyrowski},
  {Aikawa}, {Bachiller}, {Baudry}, {Benedettini}, {Bjerkeli}, {Blake},
  {Bontemps}, {Braine}, {Brinch}, {Bruderer}, {Chavarr{\'{\i}}a}, {Codella},
  {Daniel}, {de Graauw}, {Deul}, {di Giorgio}, {Dominik}, {Doty}, {Dubernet},
  {Encrenaz}, {Feuchtgruber}, {Fich}, {Frieswijk}, {Fuente}, {Giannini},
  {Goicoechea}, {Helmich}, {Herczeg}, {Jacq}, {J{\o}rgensen}, {Karska},
  {Kaufman}, {Keto}, {Larsson}, {Lefloch}, {Lis}, {Marseille}, {McCoey},
  {Melnick}, {Neufeld}, {Olberg}, {Pagani}, {Pani{\'c}}, {Parise}, {Pearson},
  {Plume}, {Risacher}, {Salter}, {Santiago-Garc{\'{\i}}a}, {Saraceno},
  {St{\"a}uber}, {van Kempen}, {Visser}, {Viti}, {Walmsley}, {Wampfler}, \&
  {Y{\i}ld{\i}z}}]{vanDishoeck2011}
{van Dishoeck}, E.~F., {Kristensen}, L.~E., {Benz}, A.~O., {et~al.} 2011,
  \pasp, 123, 138

\bibitem[{{Viti} {et~al.}(2004){Viti}, {Collings}, {Dever}, {McCoustra}, \&
  {Williams}}]{Viti2004}
{Viti}, S., {Collings}, M.~P., {Dever}, J.~W., {McCoustra}, M.~R.~S., \&
  {Williams}, D.~A. 2004, \mnras, 354, 1141

\bibitem[{{Zhang} {et~al.}(1998){Zhang}, {Ho}, \& {Ohashi}}]{Zhang1998}
{Zhang}, Q., {Ho}, P.~T.~P., \& {Ohashi}, N. 1998, \apj, 494, 636

\end{thebibliography}

\clearpage

\begin{appendix}
\section{Reference model}
\begin{table}[h!]
\caption{Column densities of species in our reference model}
\centering
\begin{tabular}{lcc}
  \hline\hline
  Molecule & G34.3+0.2 & W51/e2 \bigstrut[t] \\
 & [cm$^{-2}$] & [cm$^{-2}$] \\
 \hline
 CH$_3$OCH$_3$ & $4 \times 10^{16} $ &  $ 3 \times 10^{17}$ \bigstrut[t] \\
HCOOH t & $1 \times 10^{15}$ &  $ 1 \times 10^{16} $\\
CH$_3$CHO & $2 \times 10^{15}$ &  $1 \times 10^{16}$ \\
CH$_3$OH & $ 1 \times 10^{17}$ &  $ 3 \times 10^{18}$ \\
C$_2$H$_5$OH & $ 1 \times 10^{16}$ &  $  1 \times 10^{17}$ \\
CH$_3$CN & $ 2 \times 10^{15}$ &  $ 6 \times 10^{16}$ \\
\hline
\end{tabular}
\tablefoot{Estimated column densities of other relevant organic species. These values are used to create synthetic spectra of these molecules, which serves as a reference model when identifying lines.}
\label{t_other_coms}
\end{table}
\clearpage
\section{Observed and synthetic spectra}
\noindent\begin{minipage}{.9\textwidth}
	\centering
	\includegraphics[width=18cm]{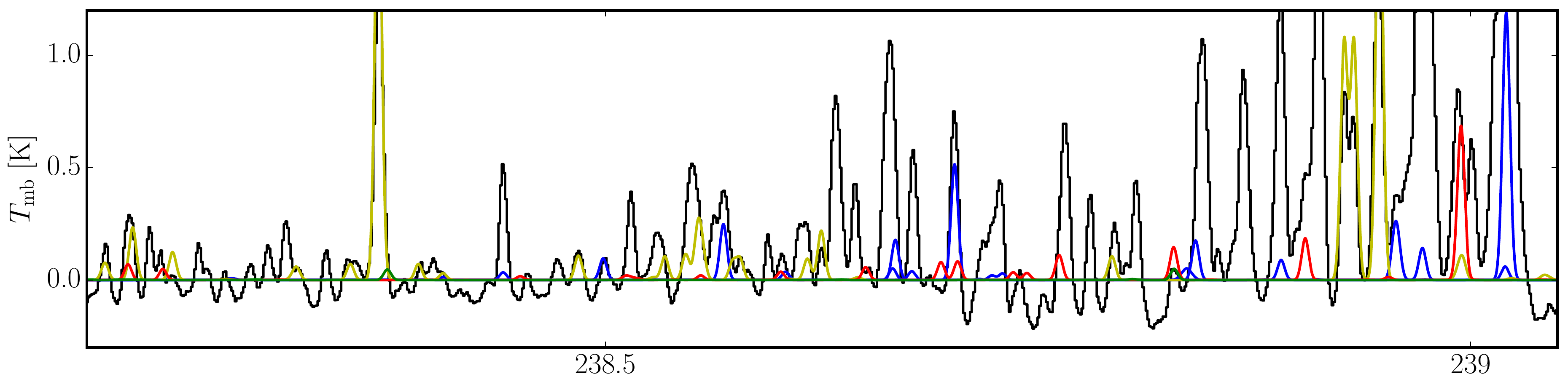}
	\includegraphics[width=18cm]{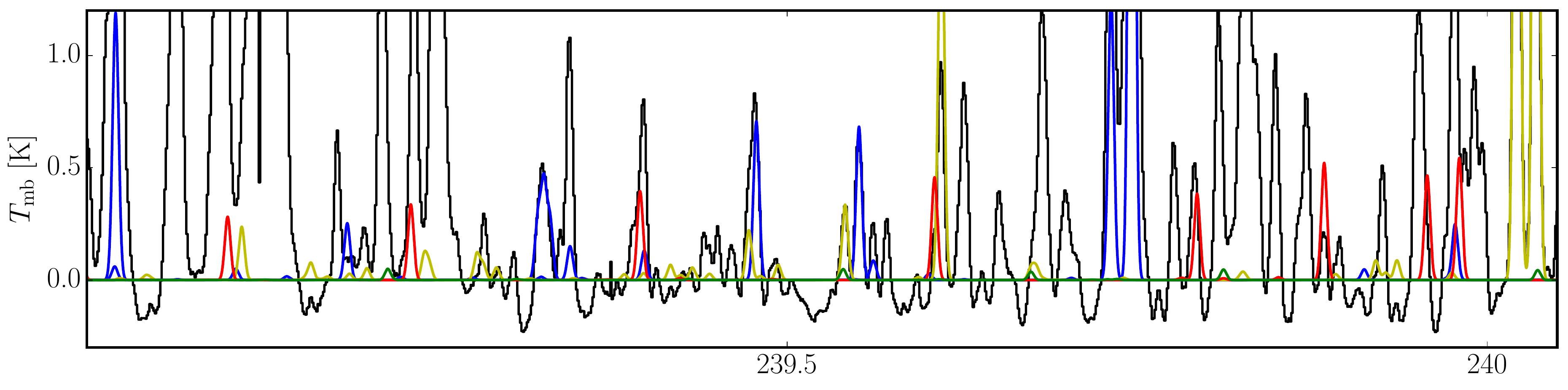}
	\includegraphics[width=18cm]{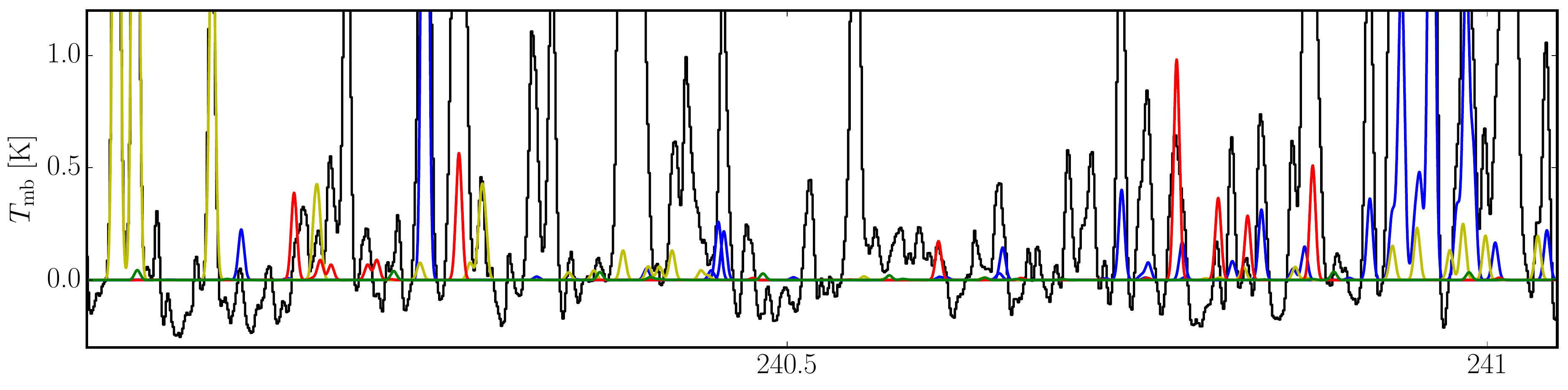}
	\includegraphics[width=18cm]{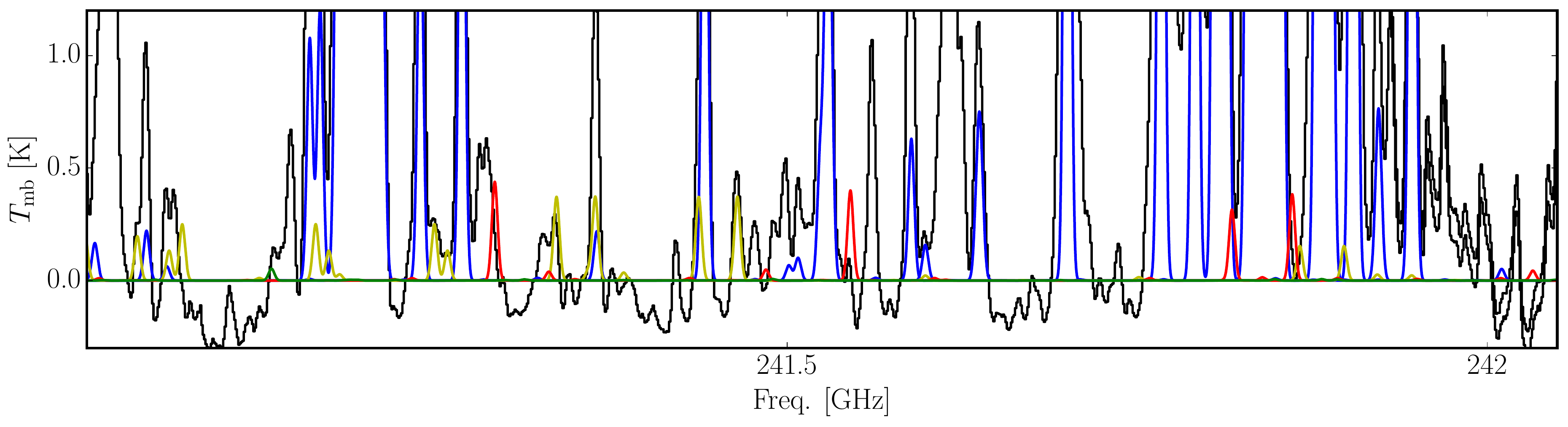}
	\captionof{figure}{Synthetic spectra (red: (CH$_2$OH)$_2$, yellow: HCOOCH$_3$, green: CH$_2$OHCHO, blue: the molecules listed in Table \ref{t_other_coms}) on top of observed spectrum (black) at 1~mm for W51/e2 in the frequency range $\sim$238-242~GHz.}
	\label{fig:long_spec_w51_1}
\end{minipage}
%
\clearpage

\noindent\begin{minipage}{.9\textwidth}
	\centering
	\includegraphics[width=18cm]{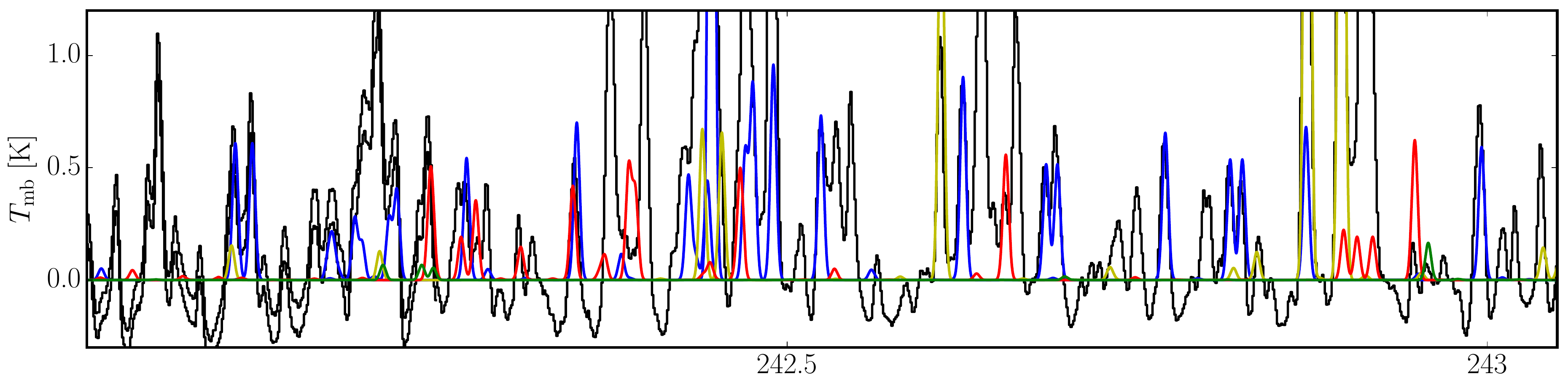}
	\includegraphics[width=18cm]{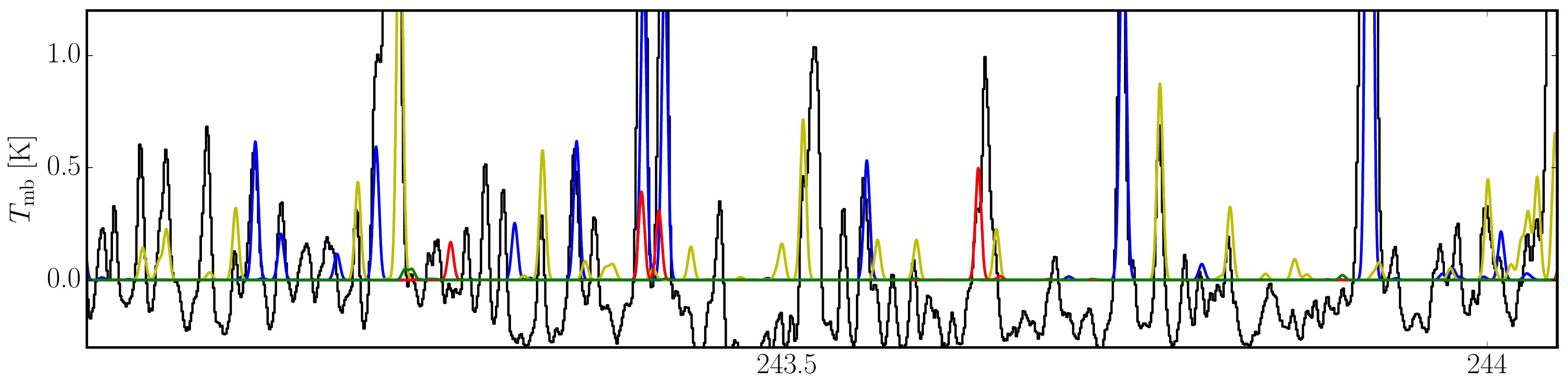}
	\includegraphics[width=18cm]{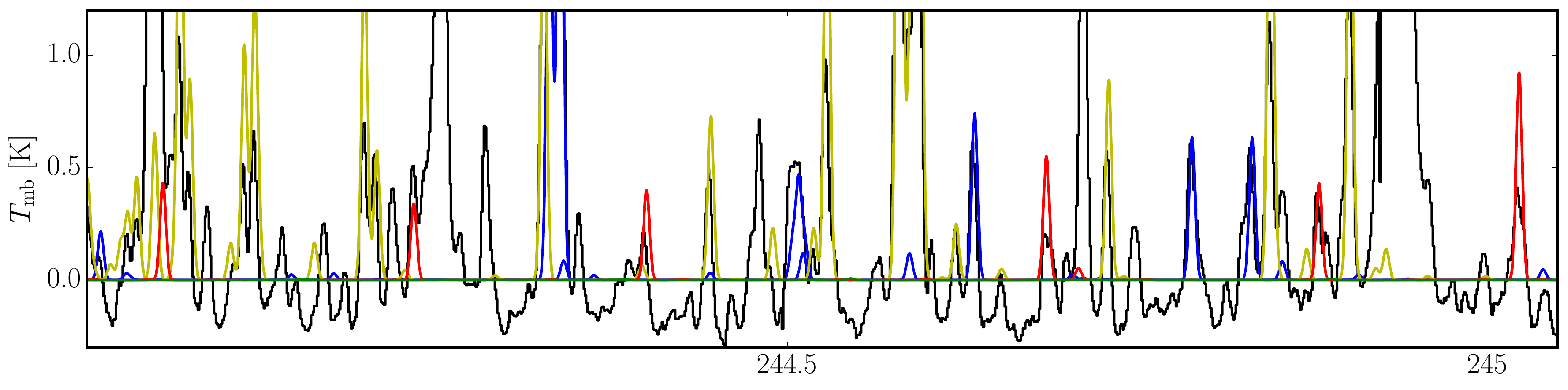}
	\includegraphics[width=18cm]{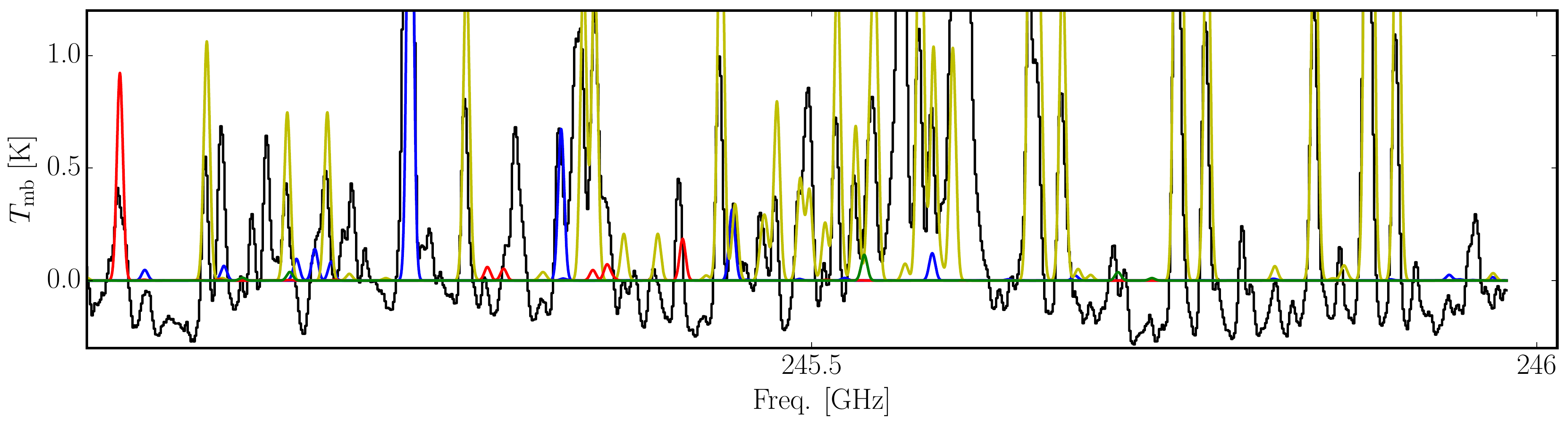}
	\captionof{figure}{Same as Figure \ref{fig:long_spec_w51_1} but for $\sim$242-246~GHz.} 
	\label{fig:long_spec_w51_2}
\end{minipage}
\clearpage
\noindent\begin{minipage}{.9\textwidth}
	\centering
	\includegraphics[width=18cm]{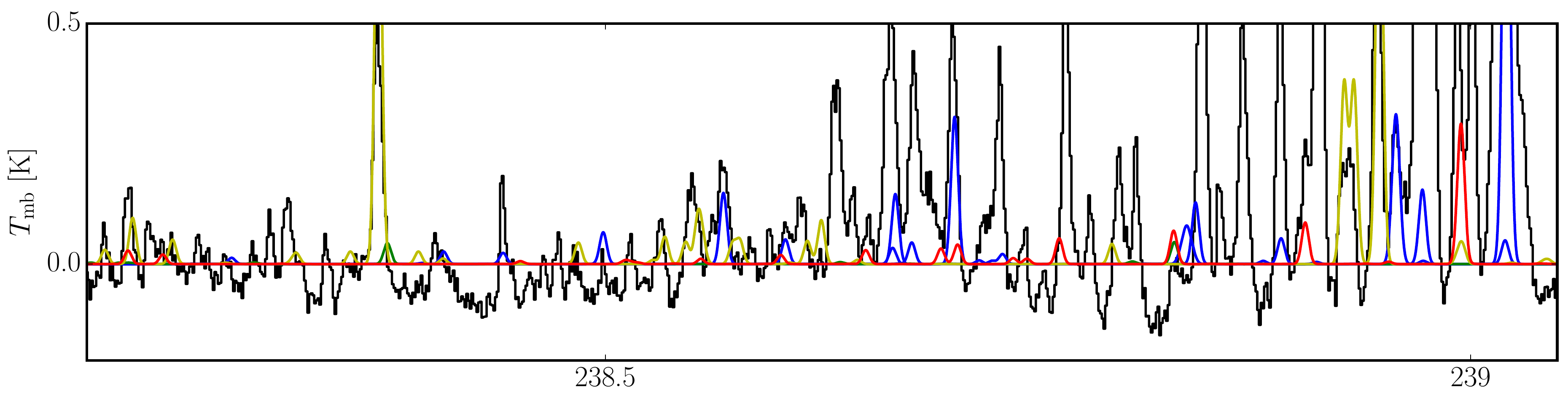}
	\includegraphics[width=18cm]{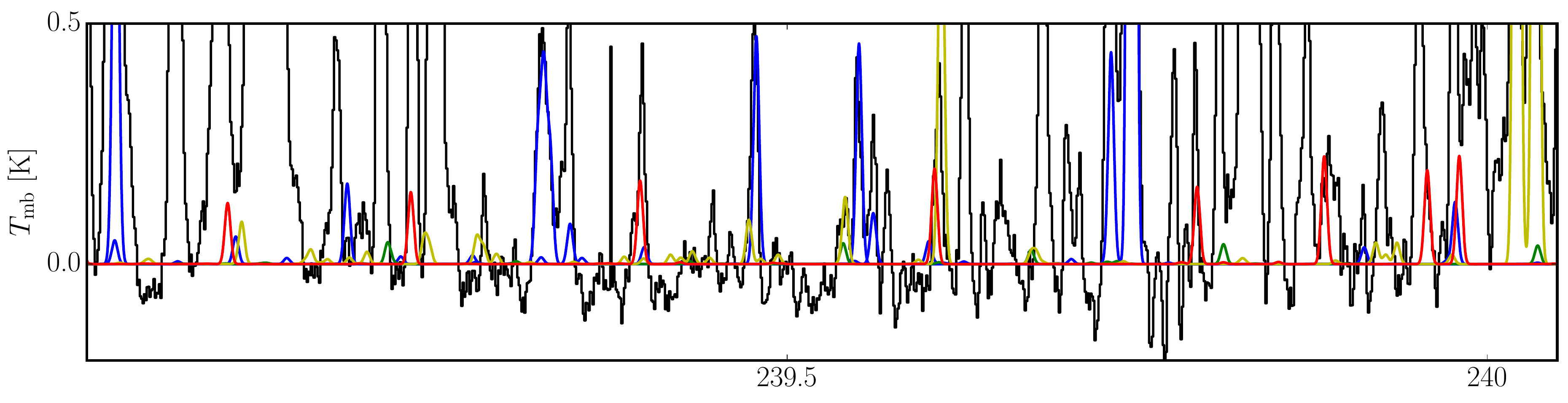}
	\includegraphics[width=18cm]{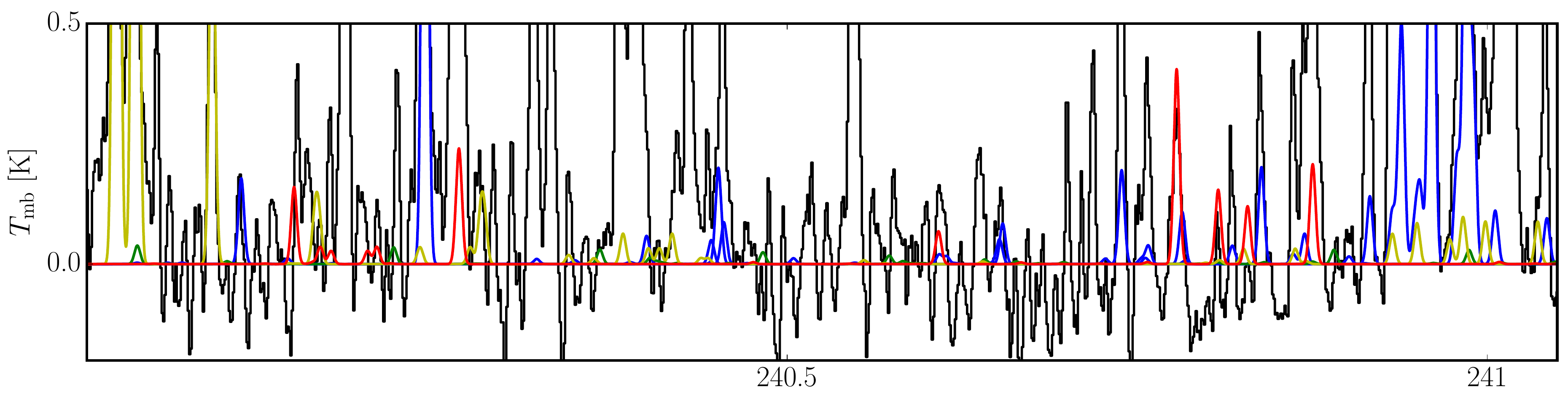}
	\includegraphics[width=18cm]{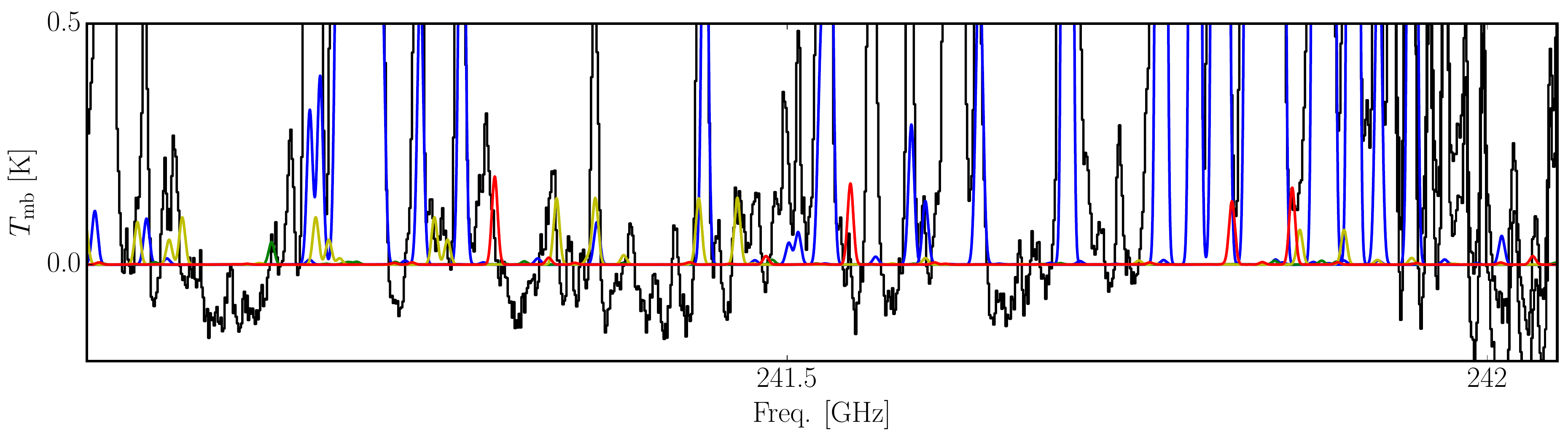}
	\captionof{figure}{Synthetic spectra (red: (CH$_2$OH)$_2$, yellow: HCOOCH$_3$, green: CH$_2$OHCHO, blue: the molecules listed in Table \ref{t_other_coms}) on top of observed spectrum (black) at 1~mm for G34.3+0.2 in the frequency range $\sim$238-242~GHz.} 
	\label{fig:long_spec_g34_1}
\end{minipage}
\clearpage
\noindent\begin{minipage}{.9\textwidth}
	\centering
	\includegraphics[width=18cm]{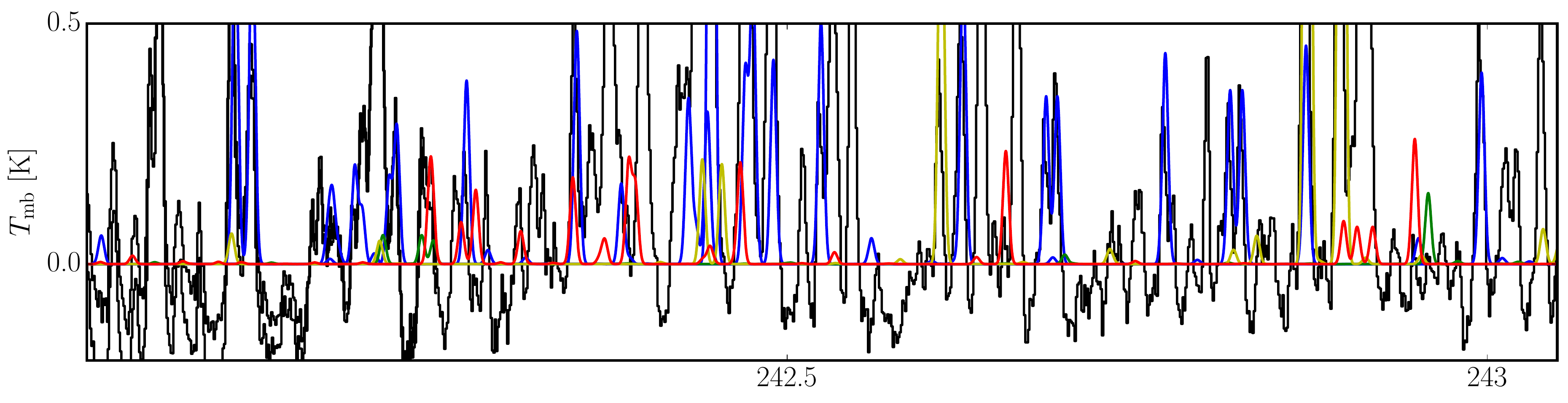}
	\includegraphics[width=18cm]{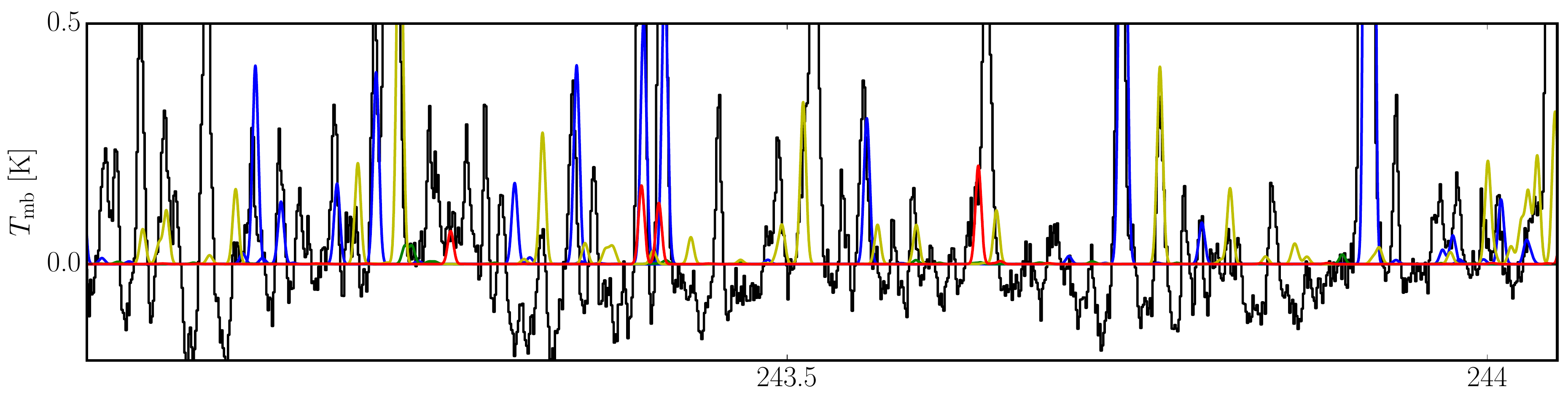}
	\includegraphics[width=18cm]{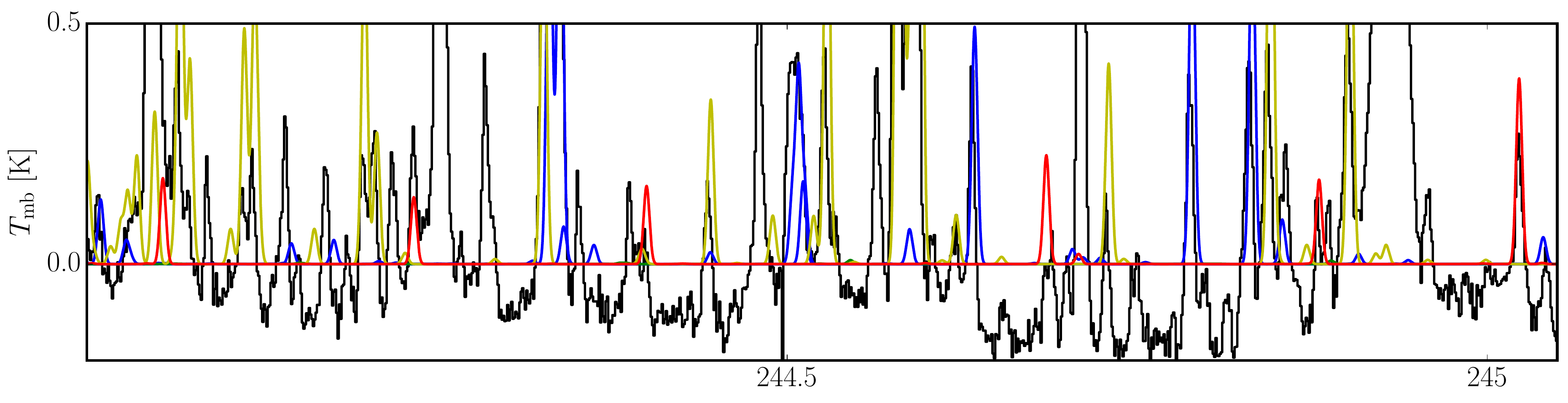}
	\includegraphics[width=18cm]{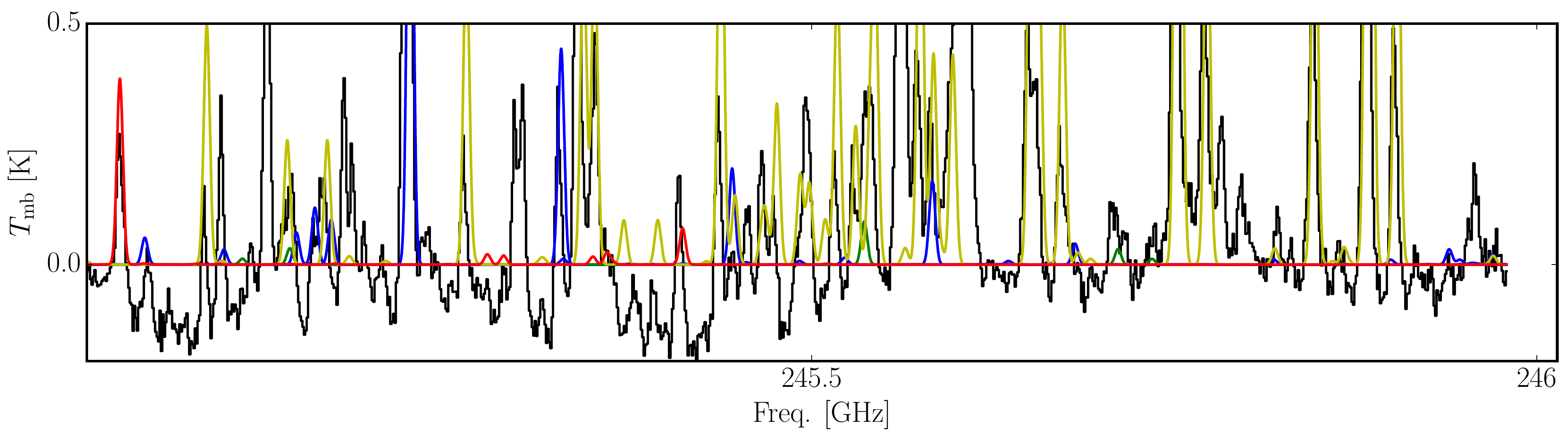}
	\captionof{figure}{Same as for Figure \ref{fig:long_spec_g34_1} but for $\sim$242-246~GHz.} 
	\label{fig:long_spec_g34_2}
\end{minipage}
\clearpage
\noindent\begin{minipage}{.9\textwidth}
	\centering
	\includegraphics[width=18cm]{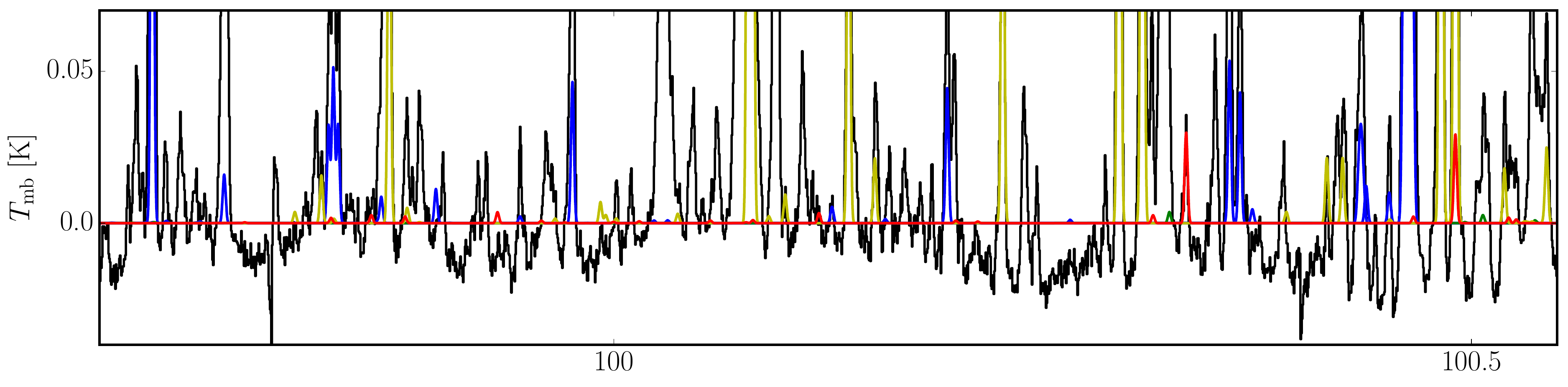}
	\includegraphics[width=18cm]{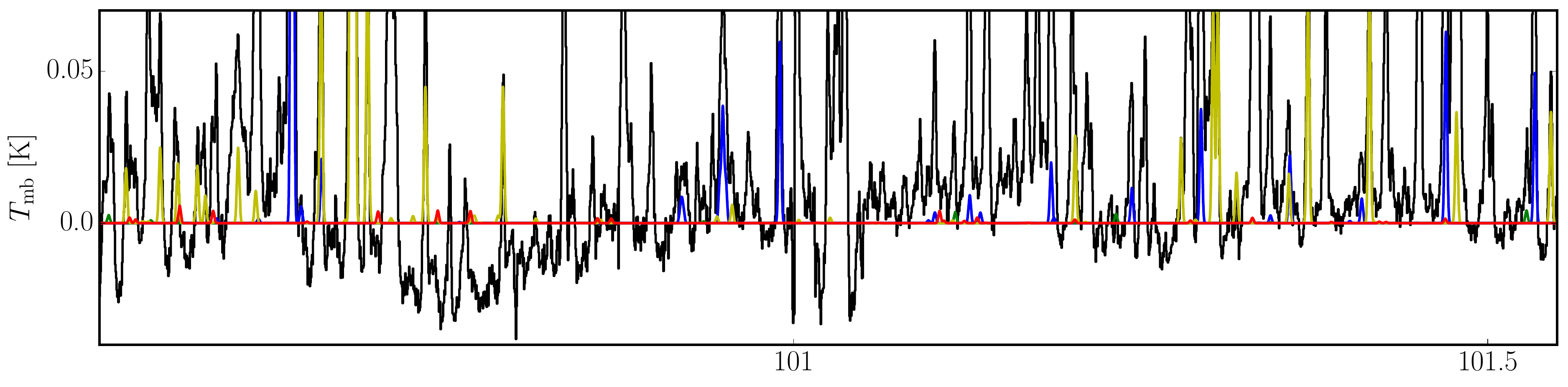}
	\includegraphics[width=18cm]{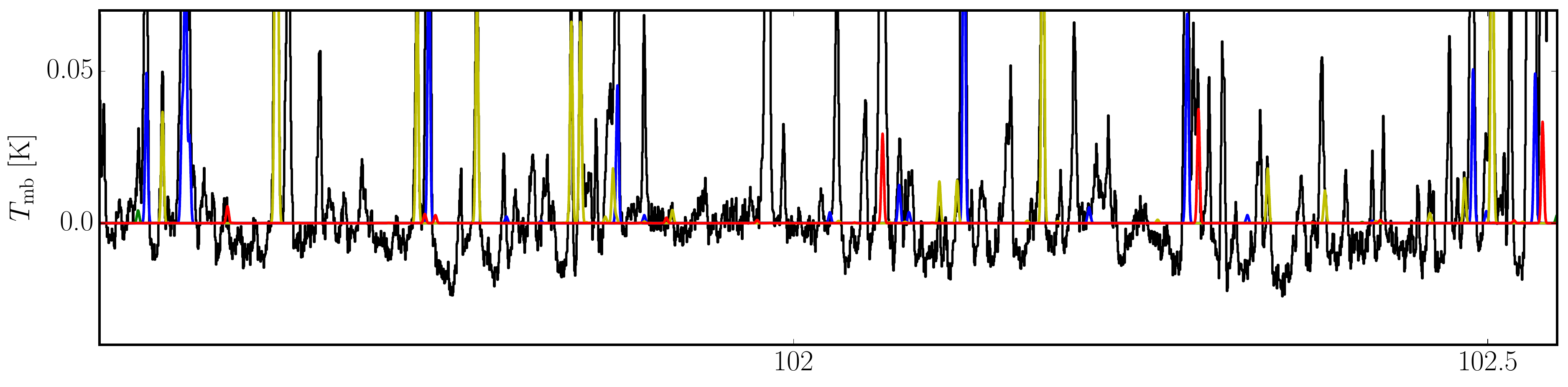}
	\includegraphics[width=18cm]{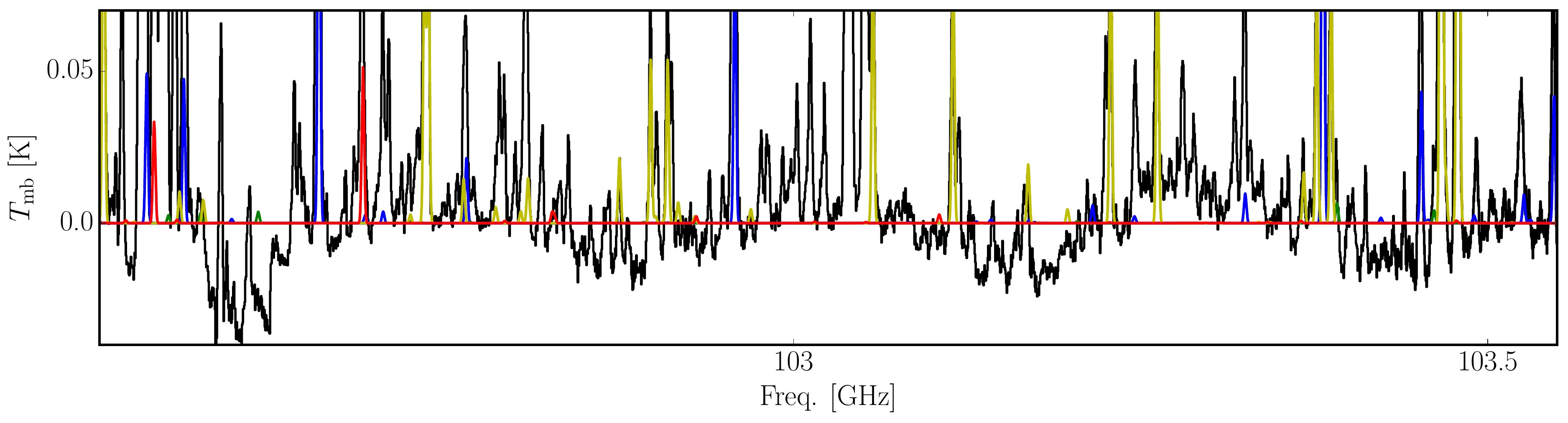}
	\captionof{figure}{Synthetic spectra (red: (CH$_2$OH)$_2$, yellow: HCOOCH$_3$, green: CH$_2$OHCHO, blue: the molecules listed in Table \ref{t_other_coms}) on top of observed spectrum (black) at 3~mm for W51/e2 in the frequency range $\sim$100-103~GHz.} 
	\label{fig:long_spec_w51_3mm_1}
\end{minipage}
\clearpage
\noindent\begin{minipage}{.9\textwidth}
	\centering
	\includegraphics[width=18cm]{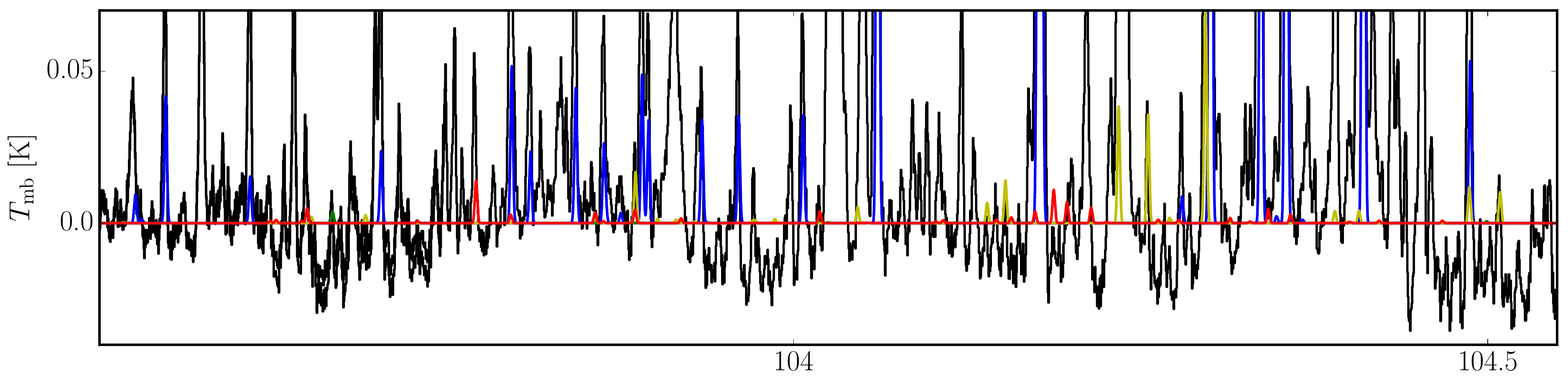}
	\includegraphics[width=18cm]{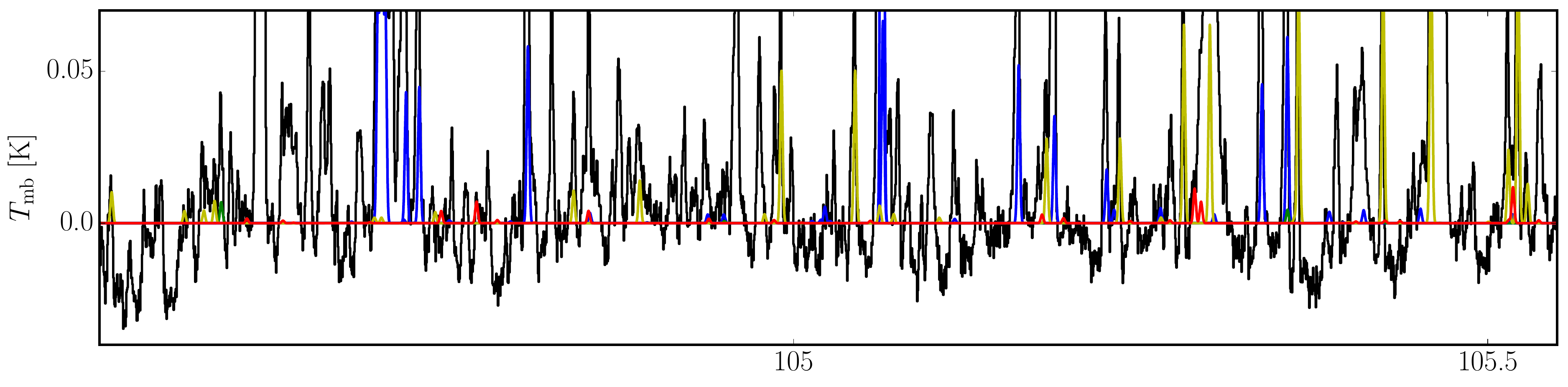}
	\includegraphics[width=18cm]{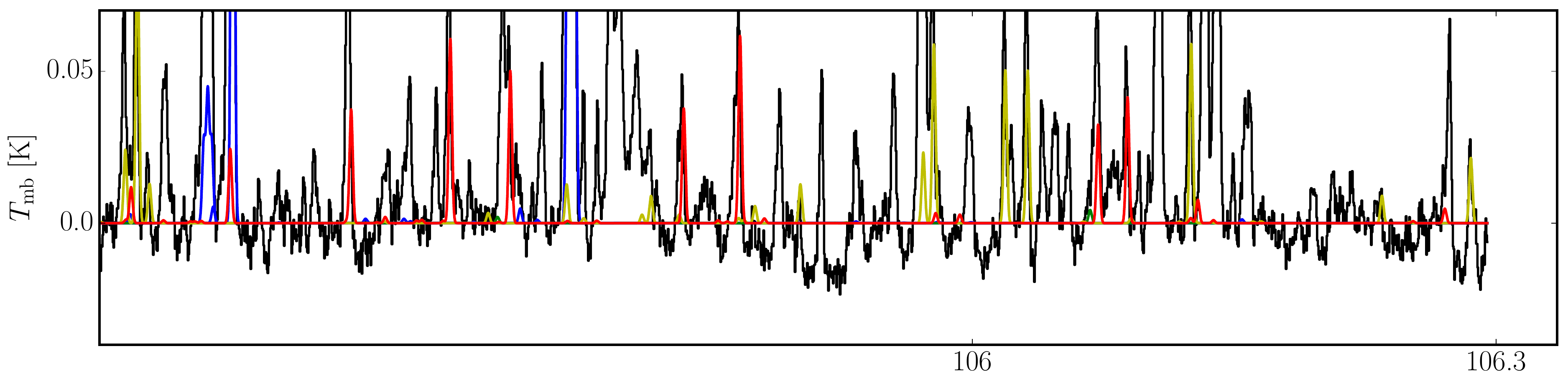}
	\captionof{figure}{Same as for Figure \ref{fig:long_spec_w51_3mm_1} but for $\sim$103-106~GHz.} 
	\label{fig:long_spec_w51_3mm_2}
\end{minipage}
\clearpage
\noindent\begin{minipage}{.9\textwidth}
	\centering
	\includegraphics[width=18cm]{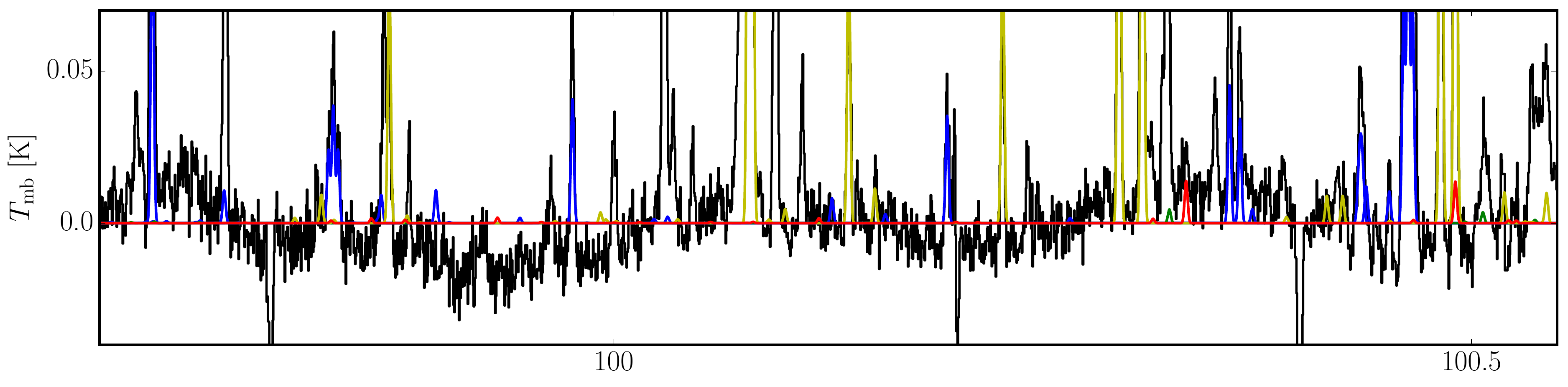}
	\includegraphics[width=18cm]{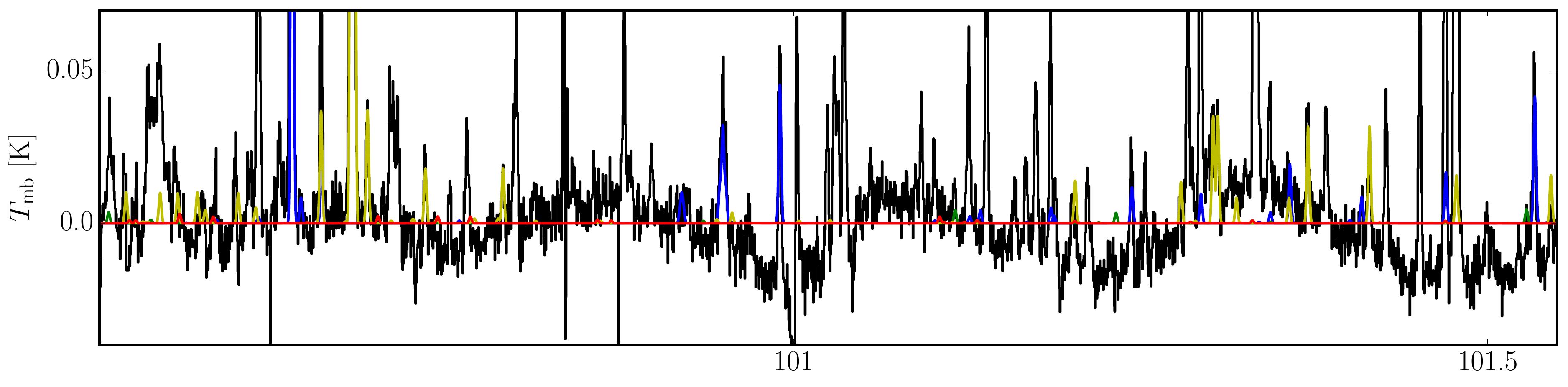}
	\includegraphics[width=18cm]{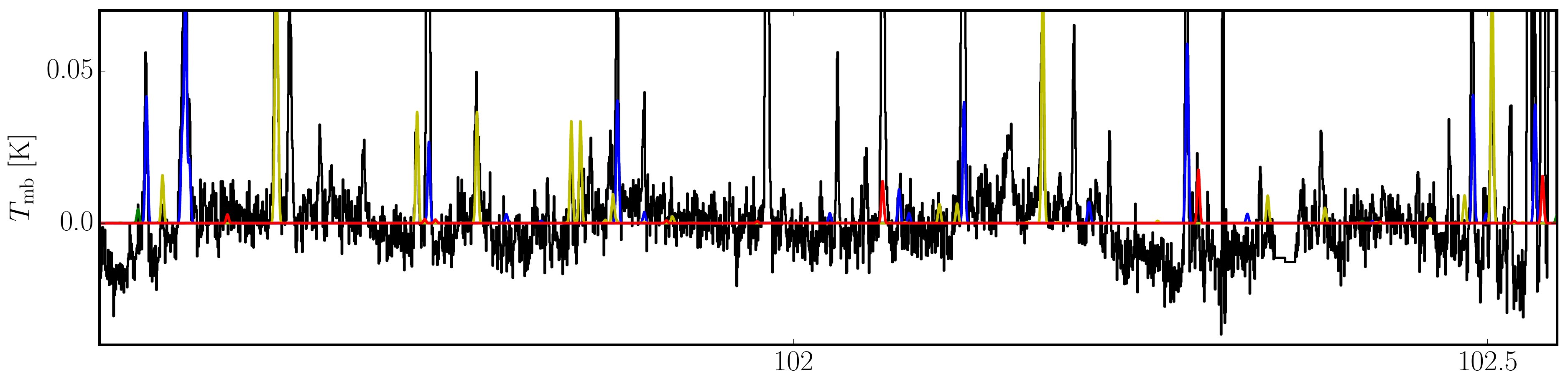}
	\includegraphics[width=18cm]{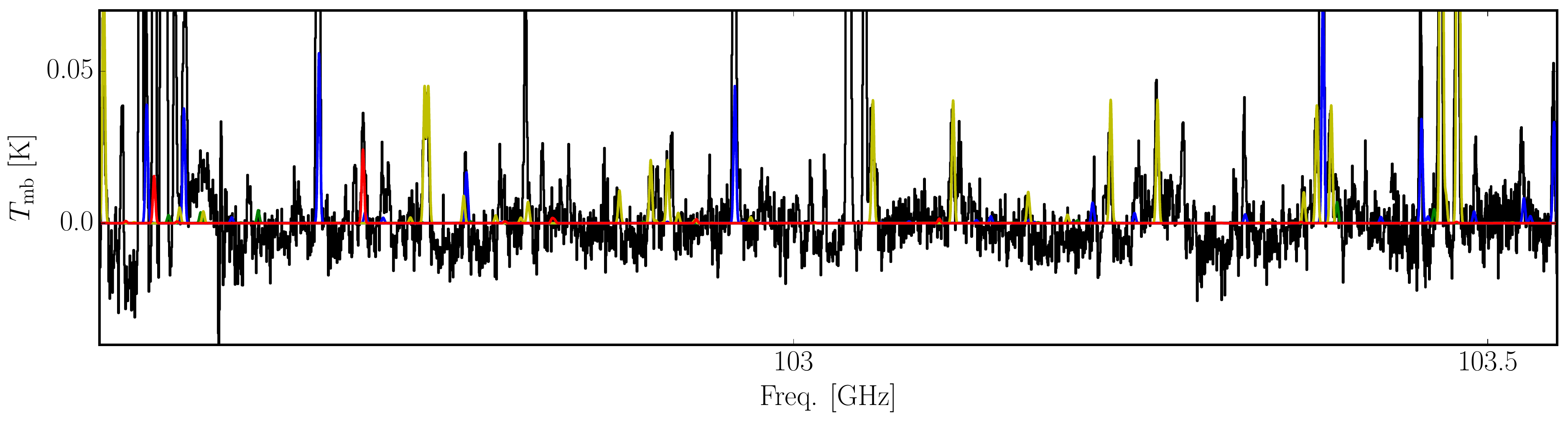}
	\captionof{figure}{Synthetic spectra (red: (CH$_2$OH)$_2$, yellow: HCOOCH$_3$, green: CH$_2$OHCHO, blue: the molecules listed in Table \ref{t_other_coms}) on top of observed spectrum (black) at 3~mm for G34.3+0.2 in the frequency range $\sim$100-103~GHz.} 
	\label{fig:long_spec_g34_3mm_1}
\end{minipage}
\clearpage
\noindent\begin{minipage}{.9\textwidth}
	\centering
	\includegraphics[width=18cm]{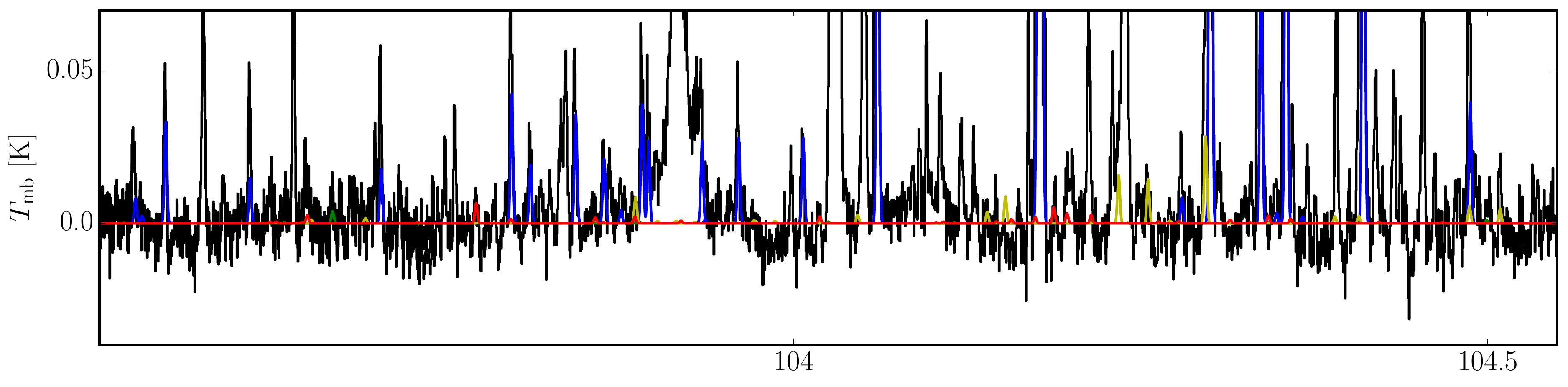}
	\includegraphics[width=18cm]{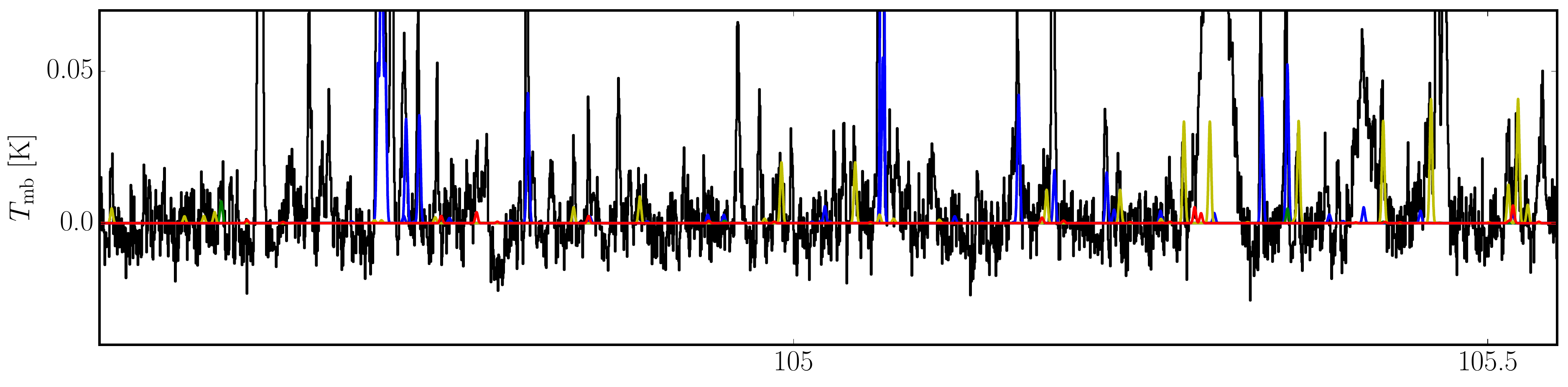}
	\includegraphics[width=18cm]{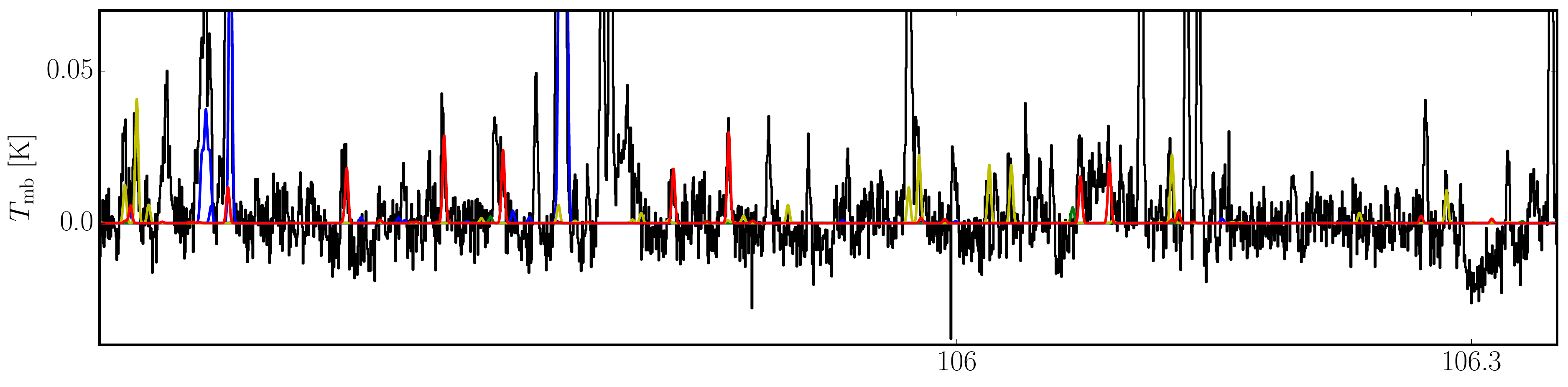}
	\captionof{figure}{Same as for Figure \ref{fig:long_spec_g34_3mm_1} but for $\sim$103-106~GHz.} 
	\label{fig:long_spec_g34_3mm_2}
\end{minipage}

\end{appendix}

\end{document}